\documentclass[preprintnumbers,eqsecnum,prd,showpacs,nofootinbib]{revtex4}
\usepackage{latexsym,amsmath,amssymb}
\usepackage[dvips]%[dvipdfm]
{graphicx,color}% Include figure files
\usepackage{dcolumn,bm}    % Align table columns on decimal point
\usepackage{subfigure}  % sub labeling of figures
\newcommand{\ma}[1]{{\mathrm{#1}}}
\newcommand{\calN}{{\cal N}}
\newcommand{\calO}{{\cal O}}
\newcommand{\calM}{{\cal M}}
\newcommand{\calC}{{\cal C}}
\newcommand{\pa}{{\partial}}
\newcommand{\frR}{{\frak R}}

\begin{document}
\thispagestyle{empty}

% \allowdisplaybreaks 

\title{Beyond $\delta N$ formalism}

\preprint{YITP-12-75}

\author{Atsushi Naruko$^{1}$}
\email[Email : ]{naruko_at_apc.univ-paris7.fr}
\author{Yu-ichi Takamizu$^{2}$}
\email[Email : ]{takamizu_at_yukawa.kyoto-u.ac.jp}
\author{Misao Sasaki$^{2}$}
\email[Email : ]{misao_at_yukawa.kyoto-u.ac.jp}

\affiliation{
$^1$ APC (CNRS-Universit\'e Paris 7), 10 rue Alice Domon et L\'eonie Duquet,
 75205 Paris Cedex 13, France \\
$^2$ Yukawa Institute for Theoretical Physics
 Kyoto University, Kyoto 606-8502, Japan
}

\date{\today}

\begin{abstract}
We develop a theory of nonlinear cosmological perturbations on
 superhorizon scales for a multi-component scalar field
 with a general kinetic term and a general form of the potential
 in the context of inflationary cosmology. 
We employ the ADM formalism and the spatial gradient expansion approach,
 characterised by $\calO(\epsilon^2)$, where $\epsilon=1/(HL)$ is
 a small parameter representing the ratio of the Hubble radius
 to the characteristic length scale $L$ of perturbations. 
We provide a formalism to obtain the solution in the multi-field case.
This formalism can be applied to the superhorizon evolution
 of a primordial non-Gaussianity beyond the so-called $\delta N$ formalism
which is equivalent to $\calO(\epsilon^0)$ of the gradient expansion. 
In doing so, we also derive fully nonlinear gauge transformation rules
valid through $\calO(\epsilon^2)$.
 These fully nonlinear gauge transformation rules can
be used to derive the solution in a desired gauge
from the one in a gauge where computations are much simpler.
As a demonstration, we consider an analytically solvable model 
 and construct the solution explicitly. 
\end{abstract}

\maketitle

%%%%%%%%%%%%%%%%%%%%%%%%%%%%%%%%%%%%%%%%%%%%%%%%%%%%%%%%%%%%%%%%%%%%%%%%%%
\section{Introduction}
\label{sec:intro}

Recent observations of the cosmic microwave background anisotropy
 show a very good agreement of the observational data with 
the predictions of conventional, single-field slow-roll models of
 inflation, that is, adiabatic Gaussian random primordial fluctuations with
 an almost scale-invariant spectrum~\cite{Komatsu:2010fb}. 
Nevertheless, possible non-Gaussianities from inflation has been a focus of 
much attention in recent years, mainly driven by recent advances
in cosmological observations.
In particular, the PLANCK satellite~\cite{Planck:2006uk} is expected to 
bring us preciser data and it is hoped that a small but finite primordial
non-Gaussianity may actually be detected.

To study possible origins of non-Gaussianity, one must go beyond 
 the linear perturbation theory. 
An observationally detectable level of non-Gaussianity
cannot be produced in the conventional, single-field slow-roll
models of inflation, since the predicted magnitude is extremely small, 
suppressed by the slow-roll parameters.
Then a variety of ways to generate a large non-Gaussianity 
 have been proposed.
 (See e.g. a focus section in CQG~\cite{CQG-focus-NG}
 and references therein for recent developments.)
They may be roughly classified into two;
 multi-field models where non-Gaussianity can be produced classically
 on superhorizon scales,
 and non-canonical kinetic term models where non-Gaussianity can be
 produced quantum mechanically on subhorizon scales. 
In particular, in the former case, the $\delta N$ formalism 
 turned out to be a powerful tool for computing non-Gaussianities
thanks to its full non-linear nature. 

On the superhorizon scales, one can employ the spatial gradient expansion 
 approach~\cite{Lifshitz:1963ps,Belinsky:1982pk,Starobinsky:1986fxa,
 Bardeen:1980kt,Salopek:1990jq,Deruelle:1994iz,Nambu:1994hu,Sasaki:1995aw,
 Sasaki:1998ug,Shibata:1999zs,Wands:2000dp,
Lyth:2004gb,Lyth:2005du,Seery:2005gb,Sasaki:2008uc,
 Tanaka:2006zp,Yokoyama:2007uu,Tanaka:2007gh,Yokoyama:2007dw,Weinberg:2008nf,
 Takamizu:2008ra,Weinberg:2008si,Takamizu:2010xy,Takamizu:2010je,
 Sugiyama:2012tj,Kodama:1997qw,Hamazaki:2008mh}.
It is characterised by an expansion parameter, $\epsilon=1/(HL)$,
 representing the ratio of the Hubble radius to the characteristic 
length scale $L$ of the perturbation. 
In the context of inflation, based on the leading order in gradient 
expansion, the $\delta N$ 
formalism~\cite{Starobinsky:1986fxa,Sasaki:1995aw,Sasaki:1998ug}
or the separate universe approach~\cite{Wands:2000dp} was developed.
It is valid when local values of the inflaton field
 at each local point (averaged over each horizon-size region) determine
 the evolution of the universe at each point. 
This leading order in the gradient expansion
 provides a general conclusion
 for the evolution on superhorizon scales that the adiabatic
growing mode is conserved on the comoving hypersurface~\cite{Lyth:2004gb}.

In this paper, we consider the curvature perturbation on superhorizon scales
up through next-to-leading order in gradient expansion, that is, 
to $\calO(\epsilon^2)$. To make our analysis as general as possible, 
we extend the $\delta N$ formalism in the following two aspects:
One is to go beyond the single-field assumption, and 
the other is to go beyond the slow-roll condition. 
While in the case of single-field inflation, the 
 curvature perturbation remains constant as mentioned above,
 the superhorizon curvature perturbation can change in time 
 in the case of multi-field inflation.
Furthermore, even for single-field inflation, 
the time evolution can be non-negligible due to a temporal violation
of the slow-roll condition. In order to study such a case, 
the $\delta N$ formalism 
 is not sufficient since the decaying mode cannot be neglected any longer,
which usually appears at $\calO(\epsilon^2)$ of gradient expansion
and is known to play a crucial role already at the level of linear perturbation
theory~\cite{Seto:1999jc,Leach:2001zf},\footnote{See, however, a
special case of single-field inflation studied recently 
in~\cite{Namjoo:2012aa} where the would-be decaying mode of the 
comoving curvature perturbation happens to be rapidly growing 
outside the horizon and an extended version of the $\delta N$ formalism 
remains to be valid although the curvature perturbation
is no longer conserved.}
not to mention the case of nonlinear perturbation 
theory~\cite{Tanaka:2006zp,Tanaka:2007gh,Takamizu:2010xy,Takamizu:2010je}.

Multi-field inflation may be motivated in the context of 
 supergravity since it suggests the existence of many flat directions in
 the scalar field potential.
In multi-field inflation, a non-slow-roll stage may appear
when there is a change in the dominating component of
the scalar field. For example, one can consider
a double inflation model in which a heavier component
dominates the first stage of inflation but damps
out when the Hubble parameter becomes smaller than the mass,
while a lighter component is negligible at the first stage
but dominates the second stage of inflation after the heavier
component has decayed out~\cite{Choi:2007su,Yokoyama:2007dw,Byrnes:2009qy}. 
However, these previous analyses are essentially based on the $\delta N$ formalism 
and it is in general necessary to extend it to $\calO(\epsilon^2)$, that is,
to the {\it beyond} $\delta N$ formalism.
We focus on the case of a multi-component scalar field.
As for a single scalar field, it has been developed in~\cite{Takamizu:2010xy}.

We mention that a multi-scalar case in the gradient expansion 
approach was studied previously~\cite{Weinberg:2008nf,Weinberg:2008si}.
However, it turns out to be valid only for a restricted situation (discussed later). 
Here we develop a general framework for fully nonlinear perturbations 
and present a formalism for obtaining the solution to $\calO(\epsilon^2)$. 
Then as an example we consider a specific model which allows an analytical
treatment of the equations of motion.

This paper is organised as follows.
In Sec.~\ref{sec:basics}, we introduce a multi-component scalar field
and derive basic equations. 
We compare several typical time-slicing conditions and
 mention the differences of them from the single-field case. 
In Sec.~\ref{sec:bdelN},
 we develop a theory of nonlinear cosmological perturbations
 on superhorizon scales. We formulate it
 on the uniform $e$-folding slicing (which is defined later). 
In Sec.~\ref{sec:sbrid}, as a demonstration of our formalism,
 we consider an analytically solvable model and give the solution explicitly. 
Sec.~\ref{sec:summary} is devoted to a summary and discussions. 
Some details are deferred to Appendices.
In Appendix~\ref{app:coincide}, the coincidence between some of time slicing
conditions is discussed by using the Einstein equations.
In Appendix~\ref{app:uniK},
 we write down the basic equations on the uniform expansion slicing,
 and study the behaviour of the curvature perturbation in this slicing.
In Appendix~\ref{app:gauge},
 we give general nonlinear gauge transformation rules valid 
 to next-to-leading order in gradient expansion. 
In Appendix~\ref{app:verify},
 we verify our formalism in a single-field model.
Finally, in Appendix~\ref{app:note}, we discuss the structure of 
the Hamiltonian and momentum constraint equations
 in the gradient expansion.

%%%%%%%%%%%%%%%%%%%%%%%%%%%%%%%%%%%%%%%%%%%%%%%%%%%%%%%%%%%%%%%%%%%%%%%%%%
\section{Basics}
\label{sec:basics}

%%%%%%%%%%%%%%%%%%%%%%%%%%%%%%%%%%%%%%%%%%%%%%%%%%%%%%%%%%%%%%%%%%%%%%%%%%
\subsection{The Einstein equations}
We develop a theory of nonlinear cosmological perturbations
 on superhorizon scales. 
For this purpose we employ the ADM formalism and
 the gradient expansion approach. 
In the ADM decomposition, the metric is expressed as 
\begin{gather}
 ds^2 = g_{\mu\nu}dx^{\mu}dx^{\nu}
 = - \alpha^2 dt^2 + \hat{\gamma}_{i j} \bigl(dx^i + \beta^i dt\bigr)
 \bigl(dx^j + \beta^j dt\bigr) \,,
\end{gather}
where $\alpha$ is the lapse function, $\beta^i$ is the shift vector and
 Latin indices run over 1,2 and 3. 
We introduce the extrinsic curvature $K_{ij}$ defined by 
\begin{align}
 K_{ij} =  \frac{1}{2 \alpha} \Bigl( \pa_t \hat{\gamma}_{ij}
 - \hat{D}_i \beta_j - \hat{D}_j \beta_i \Bigr) \,,
\label{def-Kij}
\end{align}
where $\hat{D}$ is the covariant derivative with respect to
 the spatial metric $\hat{\gamma}_{ij}$.
In addition to the standard ADM decomposition, 
 the spatial metric and the extrinsic curvature are further decomposed
 so as to separate trace and trace-free parts
\begin{align}
 \hat{\gamma}_{i j} &= a^2(t) e^{2\psi} \gamma_{i j} \,;
 \qquad \qquad \qquad \qquad \,
 \ma{det} \, \gamma_{i j} = 1 \,, \\
 K_{i j} &= a^2(t) e^{2\psi} \left( \frac{1}{3} K \gamma_{i j}
 + A_{i j} \right) \,; \qquad \gamma^{i j} A_{i j} = 0 \,,
\label{deco-Kij}
\end{align}
where $a(t)$ is the scale factor of a fiducial homogeneous 
Friedmann-Lema\^itre-Robertson-Walker (FLRW) spacetime
 and the determinant of $\gamma_{ij}$ is normalised to be unity and 
 $A_{i j}$ is trace free. 
The explicit form of $K$ is given by
\begin{gather}
 K \equiv \hat{\gamma}^{i j} K_{i j}
 = \frac{1}{\alpha} \Bigl[ 3\bigl( H + \pa_t \psi\bigr)
 - \hat{D}_i \beta^i \Bigr] \,,  
\label{def-K}
\end{gather}
 where $H$ is the Hubble parameter defined by
 $H (t) \equiv \frac{d a (t)}{d t} \Big/ a (t)$.

As for a matter field,
 let us focus on a minimally-coupled multi-component scalar field,
\begin{align}
S_m = \int d^4 x \sqrt{-g} \, P(X^{IJ},\phi^K) \,; \qquad
 X^{IJ} \equiv - g^{\mu \nu} \pa_\mu \phi^I \pa_\nu \phi^J \,,
\end{align}
 where $I$, $J$ and $K$ run over $1,2,\cdots, \calM$
with $\phi^K$ denoting the $K$-th component of the scalar field. 
Note that we do not assume a specific form of both the kinetic term 
 and potential, which are arbitrary functions of
 $X^{IJ}$ and $\phi^K$. This type of Lagrangian can be applied to, 
for example, multi-field DBI inflationary models. 
For the calculation of their non-Gaussianities, 
see, e.g.~\cite{Langlois:2008qf,Arroja:2008yy} 
and also \cite{Emery:2012sm,Kidani:2012jp} for recent developments.

The equation of motion for the scalar field is given by 
\begin{gather}
 \frac{2}{\sqrt{- g}} \pa_\mu \Bigl( \sqrt{- g} P_{( I J )} g^{\mu \nu}
 \pa_\nu \phi^J \Bigr) + P_I = 0 \,,
\label{eq-phi}
\end{gather}
 where the subscript $I$ in $P_I$ represents a derivative
 with respect to $\phi^I$ and $P_{(I J)}$ is defined as
\begin{align}
 P_{(I J)} = \frac{1}{2} \left( \frac{\pa P}{\pa X^{IJ}}
 + \frac{\pa P}{\pa X^{JI}} \right) \,.
\end{align}
The energy-momentum tensor is
\begin{align}
 T_{\mu \nu} = 2 P_{(I J)} \pa_\mu \phi^I \pa_\nu \phi^J + P g_{\mu \nu} \,.
\end{align}
Notice that this energy-momentum tensor cannot be written in 
the perfect fluid form any more, 
which is one of main differences from the single-field case. 

All the independent components of the energy-momentum tensor are
 conveniently expressed in terms of $E$ and $J_i$ as
\begin{eqnarray}
 E  \equiv  T_{\mu \nu} n^{\mu} n^{\nu} \,, \quad
 J_i  \equiv  -T_{i \mu}n^{\mu}\,,
\end{eqnarray}
and $T_{ij}$, 
 where $n^{\mu}$ is the unit vector normal to the time constant surfaces
 and is given by 
\begin{equation}
 n_{\mu}dx^{\mu} = -\alpha dt \,, \quad
  n^{\mu}\partial_{\mu} = \frac{1}{\alpha}(\partial_t-\beta^i\partial_i) \,.
\end{equation}
For convenience, we further decompose $T_{ij}$ in the same way 
as Eq.~(\ref{deco-Kij}),
 \begin{gather}
 T_{i j} = a^2 (t) e^{2 \psi} \left( \frac{1}{3} S \gamma_{i j} + S_{i j}
 \right) \,; \quad
S \equiv \gamma^{i j} T_{i j} \,.
 \end{gather}

Now we write down the Einstein equations.
In the ADM decomposition, the Einstein equations are separated into
 four constraints, the Hamiltonian constraint and three momentum constraints,
 and six dynamical equations for the spatial metric.
The constraints are
\begin{align}
 \frac{1}{a^2 e^{2\psi}} \Bigl[ R - \Bigl( 4 D^2 \psi+2D^i \psi D_i \psi \Bigr)
 \Bigr] + \frac{2}{3} K^2 - A_{i j} A^{i j} &= 2E \,, \label{basic-H}\\ 
 \frac{2}{3} \pa_i K - e^{-3\psi}D_j \Bigl( e^{3\psi} A^j{}_i \Bigr) &= J_i \,,
\label{basic-M}
\end{align}
 where $R \equiv R[\gamma]$ is the Ricci scalar of the
 normalised spatial metric $\gamma_{ij}$,
 $D_i$ is the covariant derivative with respect to $\gamma_{ij}$, 
 $D^2 \equiv \gamma^{ij} D_i D_j$,
$\gamma^{ij}$ is the inverse of $\gamma_{ij}$,
 and the spatial indices are raised or lowered by $\gamma^{ij}$
and $\gamma_{ij}$, respectively.
As for the dynamical equations for the spatial metric,
we rewrite Eq.~(\ref{def-Kij}) as
\begin{align}
 \pa_\perp \psi
 &= - \frac{H}{\alpha} 
 + \frac{1}{3} \left( K + \frac{\pa_i \beta^i}{\alpha} \right) \,, 
\label{eq-psi} \\
 \pa_\perp \gamma_{i j}
 &=  2 A_{i j} + \frac{1}{\alpha}
 \Bigl( \gamma_{i k} \pa_j \beta^k + \gamma_{j k} \pa_i \beta^k \Bigr)^{TF} \,.
\label{eq-gamma}
\end{align}
The equations for the extrinsic curvature ($K$,$A_{ij}$) are given by
\begin{align}
 \pa_\perp K &= -\frac{1}{3} K^2 - A_{i j} A^{i j}
 + \frac{1}{a^2 e^{2\psi} \alpha} \Bigl(  D^2 \alpha + D^i \alpha D_i \psi \Bigr)
 - \frac{1}{2} \left( S + E \right) \,, \\
 \pa_\perp A_{i j} &=- K A_{i j} + 2 A_i{}^k A_{k j} 
 + \frac{1}{\alpha} \left( A_{i k} \pa_j \beta^k + A_{j k} \pa_i \beta^k 
 - \frac{2}{3} A_{i j} \pa_k \beta^k \right) \notag\\
 &\qquad - \frac{1}{a^2 e^{2\psi}} \left[ R_{i j}
 + D_i \psi D_j \psi- D_i D_j \psi
 - \frac{1}{\alpha} \Bigl( D_i D_j \alpha - D_i \alpha D_j \psi 
 - D_j \psi D_i \alpha \Bigr) \right]^{TF} + S_{i j} \,,
\end{align}
where $\partial_{\perp}\equiv n^{\mu}\partial_{\mu}$,
and we have introduced the trace-free projection operator $[...]^{TF}$
defined for a tensor $Q_{i j}$ as
\begin{gather}
 Q_{i j}^{TF} \equiv Q_{i j} - \frac{1}{3} \gamma_{i j} \gamma^{k l} Q_{k l} \,. 
\end{gather}
Finally, the equations of motion for the scalar field~(\ref{eq-phi}) are
\begin{gather}
 \pa_\perp \Bigl( P_{( I J )} \pa_\perp \phi^J \Bigr)
 + K P_{( I J )} \pa_\perp \phi^J
 - \frac{1}{\alpha a^3 e^{3 \psi}} \pa_i \Bigl( \alpha a e^\psi P_{( I J )}
 \gamma^{i j} \pa_j \phi^J \Bigr) - \frac{1}{2} P_I = 0 \,.
\end{gather}

%%%%%%%%%%%%%%%%%%%%%%%%%%%%%%%%%%%%%%%%%%%%%%%%%%%%%%%%%%%%%%%%%%%%%%%%%%
\subsection{Gradient expansion and assumption}
In the gradient expansion approach we suppose that the 
 characteristic length scale $L$ of a perturbation is longer than
 the Hubble length scale $1 / H$ of the background, i.e.  $HL\gg 1$. 
Therefore, $\epsilon \equiv 1 / (HL)$ is regarded as a small parameter and
 we can systematically expand equations in the order of $\epsilon$,
 identifying a spatial derivative is of order $\epsilon$, 
$\pa_i Q = \calO (\epsilon)Q$. To clarify the order of gradient expansion,
 we introduce the superscript $(n)$.
For example, ${}^{(2)}\alpha$ means the lapse function at second order
 in gradient expansion.

As a background spacetime, we consider a FLRW universe.
At $\calO(\epsilon^0)$ of the gradient expansion,
 there is apparently no spatial gradient and the universe is 
locally homogeneous and isotropic.
This leads to the following condition on the spatial metric:
\begin{equation}
 \pa_t\gamma_{ij} = \calO(\epsilon^2) \,. 
\label{assum-gamma}
\end{equation}
Since we adopt this assumption, the spatial metric at leading order
 is given by an arbitrary spatial function of the spatial coordinates,
\begin{gather}
 {}^{(0)}\gamma_{i j} = f_{i j} (x^k) \,,
 \label{sol-gamma}
\end{gather}
under the condition that the eigenvalues of $f_{ij}$ are all positive
definite everywhere. From the definition of $A_{ij}$, Eq.~(\ref{eq-gamma}),
the above assumption implies
\begin{align}
 A_{i j} = \calO (\epsilon^2) \,.
\label{assum-Aij}
\end{align}

Throughout this paper, in order to simplify the equations,
 we set the shift vector to zero up to second order in gradient
expansion,
\begin{gather}
 \beta^i = \calO (\epsilon^3) \,. 
\end{gather}
Let us call this choice of the spatial coordinates as
 the {\it time-slice-orthogonal threading}.
Here we mention that the above condition does not completely
 fix the spatial coordinates. As discussed later,
one can actually make an arbitrary coordinate transformation
of the form, $x^i\to\bar x^i=f^i(x^j)$.

%%%%%%%%%%%%%%%%%%%%%%%%%%%%%%%%%%%%%%%%%%%%%%%%%%%%%%%%%%%%%%%%%%%%%%%%%%
\subsection{Leading order in gradient expansion}
\label{ssec:leading}

In this subsection, we study the leading order gradient expansion and
 make clear the correspondence between the leading order equations and
 background equations. This correspondence can be used to construct
the solution at leading order in gradient expansion in terms of the
background solution.

At leading order in gradient expansion, the Einstein equations are
\begin{align}
 {1\over 3}K^2 &= 2 P_{(I J)} \pa_\tau \phi^I \pa_\tau \phi^J -  P \,,
\label{BG:K} \\
  \pa_i K &= -{3} P_{(I J)} \pa_i \phi^I \pa_\tau \phi^J \,, \\
 \pa_\tau K &= -{3} P_{(I J)} \pa_\tau \phi^I \pa_\tau \phi^J \,, 
\label{BG:eq-K}
\end{align}
 and the scalar field equation is
\begin{gather}
 \pa_\tau \Bigl( P_{(I J)} \pa_\tau \phi^J \Bigr) + K P_{(I J)} \pa_\tau \phi^J
 - \frac{1}{2} P_{I} = 0 \,,
\label{BG:eq-phi}
\end{gather}
 where we have introduced the proper time $\tau$ by
\begin{gather}
 \tau (t, x^i) \equiv \int_{x^i = const.} \alpha (t', x^i) \, dt' \,.
\label{proptime}
\end{gather}
In terms of $\tau$, the expression of $K$ in Eq.~(\ref{def-K}) is simplified 
 under the time-slice-orthogonal threading condition,
 \begin{gather}
 K = \frac{1}{\alpha}
 \frac{\pa_t \bigl( a^3 e^{3 \psi} \bigr)}{a^3 e^{3\psi}} 
 = 3 \frac{\pa_\tau \bigl( a e^{\psi} \bigr)}{a e^{\psi}} \,.
\label{N:def-K}
 \end{gather}

Under the identifications,
\begin{align}
 a e^\psi \Leftrightarrow a \,, \quad \ma{and} \quad \tau \Leftrightarrow t \,,
\end{align}
 one also has the correspondence, $K \Leftrightarrow 3 H$.
This means that the basic equations at leading order,
Eqs.~(\ref{BG:eq-K}) and (\ref{BG:eq-phi}), take
exactly the same form as those in the background modulo above identifications.
Namely, given a background solution,
\begin{gather}
 \phi^I (t) \Big|_\ma{background}
 = \phi^I_\ma{BG} \Bigl[ t, \, \phi_0^I (t_0) \Bigr] \,,
\end{gather}
one can construct the solution at leading order in gradient expansion as
\begin{gather}
 \phi^I (t, x^i) \Big|_\ma{gradient}
 = \phi^I_\ma{BG} \Bigl[ \tau, \, \phi_0^I (\tau_0) \Bigr] \,.
\end{gather}
All the information of inhomogeneities is contained in the initial
condition as well as in the proper time $\tau$ through Eq.~(\ref{proptime}).
Thus it is sufficient to solve the background equations
to obtain the solution at leading order in gradient expansion.

In passing, we note that the $e$-folding number is often used as the time coordinate
 to describe the background evolution. For convenience, we define it
as the number of $e$-folds counted backward in time from a fixed final time.
That is,
\begin{gather}
 N (t) = \int_{t}^{t_0} H (t')\, dt' \,.
\end{gather}
Accordingly, the scale factor is expressed as 
\begin{align}
 a(N) = a_0 e^{- (N - N_0)} \,.
\end{align}
By replacing $t$ with $\tau$ and $H$ with $K/3$
we can generalise the $e$-fold number to 
the one defined locally in space as
\begin{gather}
 \calN (t, x^i) \equiv \frac{1}{3} \int_{t}^{t_0} d t'\,
 \alpha (t', x^i) K (t', x^i) \Bigr|_{x^i=const.} \,.
\label{def-calN}
\end{gather}
Again one can check the validity of the above correspondence by
rewriting Eqs.~(\ref{BG:K}), (\ref{BG:eq-K}) and (\ref{BG:eq-phi}) 
 in terms of $\calN$ as the time coordinate.

%%%%%%%%%%%%%%%%%%%%%%%%%%%%%%%%%%%%%%%%%%%%%%%%%%%%%%%%%%%%%%%%%%%%%%%%%%
\subsection{Various slicings and their coincidences}
\label{ssec:slice}

One needs to specify the gauge condition to study perturbations
in perturbation theory or in gradient expansion. 
Since spatial coordinates have been already fixed
 by the time-slice-orthogonal threading, 
 one has to determine the time-slicing condition. 
Here, let us list various slicings and their definitions,
\begin{align}
 \text{Comoving} ~ &; ~ J_i = 0 \,, \\
 \text{Uniform expansion} ~ &; ~ K (t, x^i) =  3 H (t) \,, \\
 \text{Uniform energy} ~ &; ~ E (t, x^i) =  E (t) \,, \\
 \text{Synchronous} ~ &; ~ \alpha (t, x^i) = 1 \,, \\
 \text{Uniform $e$-folding number} ~ &; ~ \calN (t, x^i) = N (t) \,.
\end{align}
Hereinafter, we call
the uniform expansion, uniform energy and uniform $e$-folding number
 slicings as the uniform $K$, uniform $E$ and
uniform $\calN$ slicings, respectively.

We mention that there is a remaining gauge degree of freedom
 in the synchronous or uniform $\calN$ slicing,
while the time slices are completely fixed in the uniform $K$ and 
uniform $E$ slicings.
As for the uniform $\calN$ slicing, the gauge condition demands
 $\pa_t \psi$ to vanish from Eqs.~(\ref{N:def-K}) and (\ref{def-calN}).
This means one can freely choose the initial value of $\psi$
 (and hence its spatial configuration at any later time because
$\psi$ is conserved). This corresponds to the freedom in the
choice of the initial time-slice as we see later.  
Utilising this freedom, we can make a scalar quantity,
 one of scalar fields $\phi^I$ or $K$ for example, homogeneous 
 on the initial slice.

%%%%%%%%%%%%%%%%%%%%%%%%%%%%%%%%%%%%%%%%%%%%%%%%%%%%%%%%%%%%%%%%%%%%%%%%%%
\subsection{Towards the next-to-leading order in gradient expansion}
\label{ssec:next}

As we have seen in subsection \ref{ssec:leading}, 
 the leading order solutions are given by functions of $\tau$
 in terms of the background solutions.
At next-to-leading order in gradient expansion, terms with
spatial derivatives of the leading order solution appear
in the evolution equations.
To evaluate those terms, one needs to calculate the spatial derivative of
 the lapse function, for example in $\pa_i \phi$,
 \begin{align}
 \pa_i \phi_{BG} (\tau)= \pa_\tau \phi_{BG} (\tau) \, \pa_i \tau
 = \pa_\tau \phi_{BG} (\tau) \, \int \pa_i \alpha \, dt \,.
 \end{align}
However the leading order ${}^{(0)}\alpha$ is in general given explicitly
only after solving the following equation for $\alpha$:
\begin{gather}
 \alpha = f \Bigl[ t, \, \phi (\tau) \Bigr]
 = f \left[ t, \, \phi \left( \int \alpha \, dt\right) \right] \,.
\end{gather}
As a demonstration, the analysis on the uniform $K$ slices
 is performed in Appendix~\ref{app:uniK},
and it is clearly shown that it is almost impossible to solve
 this equation, at least in an analytical way.

This problem did not appear in the single-field case.
It is because one can show that various different slicings
become identical at leading order in gradient expansion.
In particular, all the slicings listed in subsection~\ref{ssec:slice} 
 coincide with each other as shown in Appendix~\ref{app:coincide}:
\begin{gather}
 \text{comoving} = \text{uniform $K$} = \text{uniform $E$}
 ~\, \overrightarrow{\dashleftarrow} ~\, 
 \text{synchronous} = \text{uniform $\calN$} \,,
\end{gather}
 where $\rightarrow$ means the left ones imply the right ones
and $\dashleftarrow$ means it holds when one chooses the initial 
slice to be the comoving, uniform $K$ or uniform $E$ slicing 
by using the remaining gauge degree of freedom. 
Thus the lapse function is homogeneous in all the slicings
in the above, and we may set $\alpha=1$ if desired.

On the other hand, one has to face this problem
 in the case of multi-field inflation.
We overcome this problem by choosing the synchronous slicing or
 uniform $\calN$ slicing, which gives us
a {\it homogeneous time\/} coordinate.
On these slicings, one can evaluate the spatial derivatives of 
the leading order solution which appear as source terms
and construct a solution to next-to-leading order
 in gradient expansion by integrating those terms.

There are two necessary steps before reaching the goal.
Once we have a solution, it is necessary to construct a conserved
quantity out of it that can be directly related to observable quantities.
It is widely known that the comoving curvature perturbation eventually
become conserved in a single-field model in linear theory.
In non-linear theory, there exists a corresponding quantity,
 $\psi$ on the comoving, uniform $K$ or uniform $E$ slicing,
which is conserved at leading order in gradient expansion~\cite{Lyth:2004gb}.
Even in the multi-field case, the system effectively reduces
to a single-field system after the so-called non-adiabatic pressure has 
died out, that is, when the adiabatic limit is reached.
Therefore it is necessary to perform a nonlinear gauge transformation
from the uniform $\calN$ slicing to one of those three slicings. 
This is one of the steps. Since the comoving slicing is not
well defined in general in the multi-field case~\cite{Sasaki:1998ug}, 
we choose the uniform $K$ slicing as the target gauge.

The other step to be taken is related to the definition of the
{\it curvature perturbation\/} at next-to-leading order.
In linear theory the curvature perturbation is named so because 
it determines the three-dimensional Ricci scalar, and $\psi$
can be called so to full nonlinear order in the context of
the leading order in gradient expansion.
At next-to-leading order, however, $\psi$ itself is no
longer adequate to be called the curvature perturbation~\cite{Takamizu:2010xy}.
One needs to add the contribution from part of $\gamma_{i j}$, which we call 
$\chi$, to obtain a properly defined curvature perturbation 
conserved through $\calO(\epsilon^2)$. 
Therefore, after transforming from the uniform $\calN$ slicing to 
the uniform $K$ slicing, one has to evaluate the combination,
$\frR_K \equiv \psi_K + \chi_K/3$. 
This is the {\it Beyond $\delta N$ formalism}. 

Before concluding this section,
 we mention the difference between our work and that
of Weinberg~\cite{Weinberg:2008nf,Weinberg:2008si}.
There it was assumed that the lapse function can be
chosen to be equal to unity at leading order in gradient expansion,
 hence all the scalar fields are homogeneous.
This severely constrains the class of scalar field models
as well as the initial condition because the curvature perturbation 
must be always conserved at leading order in gradient expansion.
Here we do not impose such assumptions and
 perform a completely general analysis.

%%%%%%%%%%%%%%%%%%%%%%%%%%%%%%%%%%%%%%%%%%%%%%%%%%%%%%%%%%%%%%%%%%%%%%%%%%
\section{Beyond $\delta N$ formalism}
\label{sec:bdelN}

Let us first summarise the five steps in the {\it Beyond $\delta N$ formalism}.
\begin{enumerate}
 \item Write down the basic equations (the Einstein equations and
 scalar field equation) in the uniform $\calN$ slicing with
the time-slice-orthogonal threading. For convenience let us call
the choice of the coordinates in which one adopts the
 uniform $X$ slicing with the time-slice-orthogonal threading the $X$ gauge. 
So the above choice is the $\calN$ gauge.
In this gauge the metric components at leading order are trivial
since both $\psi$ and $\gamma_{ij}$ are independent of time.

\item First solve the leading order scalar field equation
 under an appropriate initial condition and then the next-to-leading order
 scalar field equation which involves spatial gradients
 of the leading order solution.

\item Solve the next-to-leading order Einstein equations for 
the metric components and their derivatives.

\item Determine the gauge transformation from the $\calN$ gauge 
 to the $K$ gauge and apply the gauge transformation rules
 to obtain the solution in the $K$ gauge. 

 \item Evaluate the curvature perturbation $\frR = \psi + \chi/3$
 in the $K$ gauge, where $\chi$ is to be extracted from $\gamma_{i j}$.
\end{enumerate}
In what follows, we describe these steps in detail but only formally.
An example in which these steps can be computed analytically will be
discussed in Sec.~\ref{sec:sbrid}.
\\

\noindent
Step 1: 
\\
First, we rewrite the uniform $\calN$ slicing condition
 from Eqs.~(\ref{N:def-K}) and (\ref{def-calN}) as
\begin{gather}
 \alpha (t, x^i) K (t, x^i) = 3 H (t)\quad \Leftrightarrow \quad
 \pa_t \psi (t, x^i) = 0 \,.
\label{N:gc}
\end{gather}
Hence $\psi$ is constant in time and is given
 by a function of the spatial coordinates alone,
\begin{gather}
 \psi (t, x^i) = \psi (x^i) \equiv C^\psi (x^i) \,.
\label{N:sol-psi}
\end{gather}

In the $\calN$ gauge, the Einstein equations are reduced to 
the following equations. 
The constraints are 
\begin{align}
 \frac{1}{a^2 e^{2 C^\psi}} \Bigl[R-\Bigl(4D^2C^\psi+2 D^iC^\psi D_iC^\psi
 \Bigr) \Bigr] + \frac{2}{3} K^2
 &= \frac{4 K^2}{9} P_{(I J)} \pa_N \phi^I \pa_N \phi^J - 2P \,,
 \label{N:Ham} \\
  \frac{2}{3} \pa_i K - e^{- 3 C^\psi} D_j \Bigl( e^{3 C^\psi} A^j{}_i \Bigr)
 &= \frac{2 K}{3} P_{(I J)} \pa_N \phi^I \pa_i \phi^J \,. 
 \label{N:Mom}
\end{align}
The evolution equations for $K$, $A_{i j}$ and $\gamma_{i j}$ are
\begin{align}
 \pa_N K
 &= K P_{(I J)} \pa_N \phi^I \pa_N \phi^J
 - \frac{3}{2 a^2 e^{2 C^\psi} K} \Bigl[ R -  \Bigl( 4D^2 C^\psi
 +2 D^i C^\psi D_i C^\psi \Bigr) \Bigr] \notag\\
 & \qquad - \frac{3}{a^2 e^{2 C^\psi}} \left[ D^2 \left( \frac{1}{K} \right)
 + D^i \left( \frac{1}{K} \right) D_i C^\psi \right]
 + \frac{3}{a^2 e^{2 C^\psi} K} P_{(I J)} D^i \phi^I D_i \phi^J \,,
 \label{N:eq-K} \\
 \pa_N A_{i j}
 &= 3 A_{i j} + \frac{3}{a^2 e^{2 C^\psi} K} \Bigl( R_{i j}
 + D_i C^\psi D_j C^\psi - D_j D_i C^\psi \Bigr)^{T F} \notag\\
 & \qquad - \frac{3}{a^2 e^{2 C^\psi}} \left[ D_i D_j \left( \frac{1}{K} \right)
 - D_i \left( \frac{1}{K} \right) D_j C^\psi - D_j \left( \frac{1}{K} \right)
 D_i C^\psi + \frac{1}{K} P_{(I J)} D_i \phi^I D_j \phi^J \right]^{T F} \,,
 \label{N:eq-Aij} \\
 \pa_N \gamma_{i j} &= - \frac{6}{K} A_{i j} \,.
 \label{N:eq-gamma} 
\end{align}
The scalar field equation is 
\begin{gather}
 \frac{K}{3} \pa_N \left( \frac{K}{3} P_{(I J)} \pa_N \phi^J \right) 
 - \frac{K^2}{3} P_{(I J)} \pa_N \phi^J
 - \frac{K}{a^2 e^{3 C^\psi}} \pa_i \left( \frac{e^{C^\psi}}{K} P_{( I J )}
 \gamma^{i j} \pa_j \phi^J \right) - \frac{1}{2} P_I = 0 \,.
 \label{N:eq-phi-pre}
\end{gather}

We rewrite Eq.~(\ref{N:eq-phi-pre}) by eliminating $\pa_N K$
 with Eq.~(\ref{N:eq-K}) as
\begin{align}
 & \frac{K^2}{9} \pa_N\Bigl(P_{(IJ)} \pa_N \phi^J\Bigr)
 + \frac{K^2}{9} \Bigl( P_{(K L)} \pa_N \phi^K \pa_N \phi^L - 3 \Bigr)
 P_{(IJ)} \pa_N \phi^J - {1\over 2}P_I \notag\\
 & \qquad
 = \frac{K}{a^2 e^{2 C^\psi}} \left\{ \frac{1}{e^{C^\psi}} \pa_i \left(
 \frac{e^{C^\psi}}{K} P_{(I J)} \gamma^{i j} \pa_j \phi^J \right) 
 + \frac{1}{3} \left[ D^2 \left( \frac{1}{K} \right) 
 + D^i \left( \frac{1}{K} \right) D_i C^\psi \right] P_{(I J)}\pa_N \phi^J
 \right\} \notag\\
 & \qquad \qquad
 + \frac{1}{6 a^2 e^{2 C^\psi}} \left[ R - \Bigl( 4 D^2 C^\psi
 + 2 D^i C^\psi D_i C^\psi \Bigr) - 2 P_{(K L)} \gamma^{i j}
 \pa_i \phi^K \pa_j \phi^L \right] P_{(IJ)}\pa_N \phi^J \,.
\label{N:eq-phi}
\end{align}
Thus once $K^2$ is expressed in terms of the scalar field and its derivatives,
 the above equation gives a closed scalar field equation.
An explicit derivation of the closed scalar field equation is
possible only after we specify the explicit form of
 $P(X^{IJ},\phi)$ as a function of $X^{I J}$ and $\phi$.
Here we describe generally but formally the procedure to obtain $K^2$ 
as a function of the scalar field.

First, we separate the term with time-derivatives in $X^{IJ}$
and denote it by $K^2 Y^{I J}$ where $Y^{IJ}\equiv \pa_N\phi^I \pa_N\phi^J/9$, 
\begin{align}
 X^{I J} = K^2 Y^{I J}
 - \frac{1}{a^2 e^{2 C^\psi}} \gamma^{i j} \pa_i \phi^I\pa_j \phi^J \,.
\label{def-Yij}
\end{align}
At leading order, we can neglect the spatial-derivative term
 in Eq.~(\ref{def-Yij}).
Then, $P$ and $P_{(IJ)}$ are given by $P(K^2 Y^{IJ},\phi^I)$
 and $P_{(IJ)}(K^2 Y^{IJ},\phi^I)$, respectively. 
To next-to-leading order, $P$ and $P_{(IJ)}$ can be expanded as 
\begin{align}
 P (X^{I J}, \phi^K)
 &= P (K^2 Y^{I J},\phi^K)
 - \frac{1}{a^2 e^{2 C^\psi}} \gamma^{i j} 
\pa_i {}^{(0)}\phi^{I} \pa_j {}^{(0)}\phi^{J}
 \frac{\pa P}{\pa X^{I J}} + \calO (\epsilon^4) \,, \\
 P_{(I J)} (X^{K L}, \phi^M)
 &= P_{(I J)} (K^2 Y^{K L},\phi^M)
 - \frac{1}{a^2 e^{2 C^\psi}} \gamma^{i j}
 \pa_i {}^{(0)}\phi^{K}\pa_j {}^{(0)}\phi^{L}
 \frac{\pa P_{(I J)}}{\pa X^{K L}} + \calO(\epsilon^4) \,.
\end{align} 
Inserting these expressions into Eq.~(\ref{N:Ham}),
 one obtains an algebraic equation for $K^2$. Solving it
gives an expression of $K^2$ in terms of the scalar field.
Then a closed equation for the scalar field 
is obtained by plugging it into Eq.~(\ref{N:eq-phi}).
\\

\noindent
Step 2:
\\
Although we can keep our discussion completely general,
below we focus on the case of a multi-component canonical scalar field,
\begin{align}
 P = \frac{1}{2} \delta_{I J} X^{I J} - V (\phi^1, \cdots, \phi^\calM)
 \quad \ma{and} \quad
 P_{(I J)} = \frac{1}{2} \delta_{I J} \,.   
 \end{align}
This choice is taken purely for the sake of simplicity and
clarity, because the expansion $K$ can be explicitly expressed in terms of 
the scalar field in this case. In general, one cannot obtain an explicit
expression of $K$ in terms of the scalar field
unless the form of $P$ is explicitly specified.
Nevertheless, the discussion below also applies to the general case 
perfectly.

 From Eq.~(\ref{N:Ham}) we find
\begin{gather}
 \frac{K^2}{9} 
 = \left( 1 - \frac{1}{6} \pa_N \phi_I \pa_N \phi^I \right)^{-1} \left\{
\frac{V}{3} - \frac{1}{6 a^2 e^{2 C^\psi}} \left[ R -  \Bigl(4 D^2 C^\psi
 +2 D^i C^\psi D_i C^\psi \Bigr) + D^i \phi_I D_i \phi^I \right]
 \right\} \,. 
\label{N:Ham-phi}
\end{gather}
Inserting this into Eq.~(\ref{N:eq-phi}),
 one obtains the following closed equation:
\begin{align}
 & \left( 1 - \frac{1}{6} \pa_N \phi_J \pa_N \phi^J \right)^{-1}
 \pa_N^2 \phi_I - 3 \pa_N \phi_I + 3 \frac{V_I}{V} \notag\\
 & \quad = \frac{K}{a^2 e^{2 C^\psi} V} \left\{ \frac{3}{e^{C^\psi}} \pa_i
 \left( \frac{e^{C^\psi}}{K} \gamma^{i j} \pa_j \phi_I \right) 
 + \left[ D^2 \left( \frac{1}{K} \right)
 + D^i \left( \frac{1}{K} \right) D_i C^\psi \right] \pa_N \phi_I \right\}
 - \frac{D^i \phi_J D_i \phi^J}{a^2 e^{2 C^\psi} V} \pa_N \phi_I \notag\\
 & \qquad
 + \frac{1}{2 a^2 e^{2 C^\psi} V} \left[ R
 - \Bigl(4 D^2 C^\psi + 2D^i C^\psi D_i C^\psi \Bigr)
 + D^i \phi_J D_i \phi^J \right]
 \left[ \left( 1 - \frac{1}{6} \pa_N \phi_K \pa_N \phi^K \right)^{-1}
 \pa_N^2 \phi_I - 2 \pa_N \phi_I \right] \,.
\label{N:closed-phi}
\end{align}
Since each term in the right-hand side of the equation
involves two spatial derivatives, we can neglect them at leading order.
At next-to-leading order, they can be understood as source terms 
 whose time evolution have been already determined
from the leading order solution. 
As noted in the above, although one cannot obtain an equation
like the above explicitly for general $P$, one can always derive
a closed equation for the scalar field once a specific form of $P$
is given.

Solving the closed equation for the scalar field, the solution
is formally obtained as
\begin{eqnarray}
{}^{(0)}\phi_\calN&=&{}^{(0)}\phi_\calN
\Bigl[N; {}^{(0)}\phi_0, {}^{(0)}\pa_N \phi_0, 
C^\psi, {}^{(0)}\gamma_{i j 0} 
\Bigr]\,,
\cr
{}^{(2)}\phi_\calN&=&{}^{(2)}\phi_\calN
\Bigl[ N; {}^{(2)}\phi_0, 
{}^{(2)}\pa_N \phi_0, D({}^{(0)}\phi_0, 
{}^{(0)}\pa_N\phi_0, C^\psi, {}^{(0)}\gamma_{i j 0}) 
\Bigr]\,,
\label{scalarsol}
\end{eqnarray}
where the subscript $\calN$ indicates the $\calN$ gauge.
In the arguments on the right-hand side, the subscript $0$ denotes the 
initial value, and $D(\cdots)$ the spatial derivatives of the quantities
inside the parentheses.
\\

\noindent
Step 3:
\\
Once the solution for the scalar field is obtained, 
 one can solve Eqs.~(\ref{N:eq-K}), (\ref{N:eq-Aij}) and
 (\ref{N:eq-gamma}) to obtain the metric quantities. 
As for $K$, however, it is simpler and in fact better to use
Eq.~(\ref{N:Ham-phi}), which is essentially the Hamiltonian constraint,
since the integral constants appearing from integrating Eq.~(\ref{N:eq-K})
are not freely specifiable but must satisfy the Hamiltonian constraint.
\\

\noindent
Step 4:
\\
Given the solution in the $\calN$ gauge, the next step is to find
the gauge transformation from the $\calN$ gauge to the $K$ gauge. 
It can be determined as follows. As noted above, the expression
for $K_\calN$ in terms of the scalar field is obtained from
Eq.~(\ref{N:Ham-phi}). Since the leading order and next-to-leading order
scalar field solutions are expressed as Eqs.~(\ref{scalarsol}), 
the same is true for $K_\calN$,
\begin{align}
 &{}^{(0)}K_\calN
 = {}^{(0)}K_\calN\Bigl[ N; {}^{(0)}\phi_0, {}^{(0)}\pa_N \phi_0, 
C^\psi, {}^{(0)}\gamma_{i j 0} \Bigr]\,,
\notag\\
 &{}^{(2)}K_\calN
 = {}^{(2)}K_\calN\Bigl[ N; {}^{(2)}\phi_0, 
{}^{(2)}\pa_N \phi_0, D({}^{(0)}\phi_0, 
{}^{(0)}\pa_N \phi_0, C^\psi, {}^{(0)}\gamma_{i j 0}) \Bigr]\,.
\label{N:sol-K}
\end{align}

Let the transformation from the uniform $\calN$ slicing to the
uniform $K$ slicing be given by
$N \to \tilde{N} = N + n (N, x^i)$ or conversely
 $\tilde{N} + \tilde{n} (\tilde{N}, \tilde{x}^i) = N$,
where the uniform $K$ slice is given by $\tilde{K}=$const..
This nonlinear  gauge transformation is discussed in detail in
 Appendix~\ref{app:gauge}. 
The nonlinear gauge transformation generator $\tilde{n}(\tilde{N},x^i)$ 
from the $\calN$ gauge to $K$ gauge is then determined by
the condition that $K$ is spatially homogeneous on the uniform $K$
slice by definition:
\begin{align}
&{}^{(0)}K_K (\tilde{N},\tilde{x}^i)={}^{(0)}K_{\calN} 
(\tilde{N}+{}^{(0)}\tilde{n}, x^i)=0\,,
\label{N:step4-K}\\
&{}^{(2)}K_K (\tilde{N},\tilde{x}^i)={}^{(2)}K_K
\Bigl[ \tilde{N}+{}^{(2)}\tilde{n}, D({}^{(0)}K_{\calN},{}^{(0)}\tilde{n}) 
\Bigr]=0\,.
\label{N:step4-K2}
\end{align}
Then the other quantities in the $K$ gauge are obtained by applying
the above gauge transformation. In particular, the determinant of the
spatial metric $\psi_K$ in the $K$ gauge is obtained from 
Eqs.~(\ref{GT-psi-0}) and (\ref{GT-psi-2}), and the unimodular part of
the spatial metric $\gamma_{ij\,K}$ from 
Eqs.~(\ref{GT-gamma-0}) and (\ref{GT-gamma-2}). 
\\

\noindent
Step 5:
\\
We are to construct the nonlinear curvature perturbation,
$\frR_K=\psi_K+\chi_K/3$, where $\chi_K$ is the scalar part of the
metric $\gamma_{ij\,K}$. The scalar part of $\gamma_{ij}$ is defined
in the same way as the single-scalar case~\cite{Takamizu:2008ra}, 
\begin{align}
 \chi \equiv - \frac{3} {4}  \triangle^{-1}  
\Bigl\{\pa^i e^{-3\psi} \pa^j \Bigl[e^{3\psi} \bigl( \gamma_{ij}
 - \delta_{ij} \bigr) \Bigr] \Bigr\}\,,
\label{sb:def-chi}
\end{align}
where $\triangle^{-1}$ is the inverse Laplacian operator on the flat background.
Extracting $\chi_K$ from $\gamma_{ij\,K}$, and combining $\psi_K$ and $\chi_K$,
we finally obtain the nonlinear curvature perturbation
$\frR_K=\psi_K+\chi_K/3$ in the $K$ gauge.

%%%%%%%%%%%%%%%%%%%%%%%%%%%%%%%%%%%%%%%%%%%%%%%%%%%%%%%%%%%%%%%%%%%%%%%%%%
\section{solvable example}
\label{sec:sbrid}
In this section, we demonstrate how to obtain the solution up to
next-to-leading order in gradient expansion by applying our formalism
to a specific, analytically solvable model.

%%%%%%%%%%%%%%%%%%%%%%%%%%%%%%%%%%%%%%%%%%%%%%%%%%%%%%%%%%%%%%%%%%%%%%%%%%
\subsection{Model and equations}
For simplicity, we consider a canonical scalar field 
 with exponential potential~\cite{Sasaki:2008uc}, 
 \begin{align}
 P = \frac{1}{2} X_{I J} - V (\phi_I) \,, \qquad 
 V (\phi_I) = W \exp \left[ \sum_J m_J \phi_J
 \right] \,,
 \end{align}
 where $W$ is a constant.
The leading order scalar field equation is given by
\begin{gather}
 \left( 1 - \frac{1}{6} \pa_N {}^{(0)}\phi_J \pa_N {}^{(0)}\phi_J\right)^{-1}
 \pa_N^2 {}^{(0)}\phi_I - 3 \pa_N {}^{(0)}\phi_I + 3m_I = 0 \,,
\label{sb:eq-phi-0}
\end{gather}
 where we have omitted a summation symbol over the
field component indices, $J$. Hereafter summation is implied
over repeated component indices.

Further we assume the two slow-roll conditions on the leading order equation.  
The first one is that we can neglect the ``kinetic energy''
 in the energy density of the scalar field,
\begin{align}
 \pa_N {}^{(0)}\phi_I \pa_N {}^{(0)}\phi_I \ll 1 \,.
\label{slow-roll-cond}
\end{align}
The second one is that we can neglect the ``acceleration'',
\begin{align}
 \pa_N^2 {}^{(0)}\phi_I\ll \pa_N {}^{(0)}\phi_I\,.
\label{slow-roll-cond2}
\end{align}
It is important that we apply these assumptions only to ${}^{(0)}\phi$.
We do not impose the slow-roll conditions on ${}^{(2)}\phi$.
So we can rewrite Eq.~(\ref{sb:eq-phi-0}) as
\begin{gather}
 \pa_N {}^{(0)}\phi_I - m_I = 0 \,.
\label{sb:eq-phi-0-slow}
\end{gather}
At next-to-leading order, we have
\begin{align}  
 \pa_N^2 {}^{(2)}\phi_I - 3 \pa_N {}^{(2)}\phi_I
 + (\pa_N {}^{(0)}\phi_I - m_I) \pa_N{}^{(0)}\phi_J \pa_N{}^{(2)}\phi_J
 = \frac{S_I^\phi}{a^2 e^{2 C^\psi} {}^{(0)}V} \,,
\label{sb:eq-phi-2-slow-pre}
\end{align}
 where
\begin{align}
 S^\phi_I 
 &= 3 \frac{{}^{(0)}K}{e^{C^\psi}} \pa_i \left( \frac{e^{C^\psi}}{{}^{(0)}K}
 D^i{}^{(0)}\phi_I \right) + {}^{(0)}K \left[ D^2 \left( \frac{1}{{}^{(0)}K} \right)
 + D^i \left( \frac{1}{{}^{(0)}K} \right) D_i C^\psi \right] \pa_N {}^{(0)}\phi_I
 \notag\\
 & \qquad
 - \Bigl[ R - \Bigl( 4D^2 C^\psi + 2 D^i C^\psi D_i C^\psi \Bigr)
 + 2 D^i{}^{(0)}\phi_J D_i {}^{(0)}\phi_J \Bigr] \pa_N {}^{(0)}\phi_I \,.
\label{sb:sphi}
\end{align}
Inserting the leading order equation (\ref{sb:eq-phi-0-slow}) 
 into (\ref{sb:eq-phi-2-slow-pre}), it gives 
\begin{align}  
 e^{3N} \pa_N \Bigl( e^{- 3 N} \pa_N {}^{(2)}\phi_I \Bigr)
 = \frac{S_I^\phi}{a^2 e^{2 C^\psi} {}^{(0)}V} \,.
\label{sb:eq-phi-2-slow}
\end{align}
This is the basic equation at next-to-leading order.

%%%%%%%%%%%%%%%%%%%%%%%%%%%%%%%%%%%%%%%%%%%%%%%%%%%%%%%%%%%%%%%%%%%%%%%%%%
\subsection{Solution}
We can easily solve the leading order and next-to-leading order
scalar field equations.
The solution of Eq.~(\ref{sb:eq-phi-0-slow}) is given by
\begin{align}
 {}^{(0)}\phi_I (N) = C_I^\phi + m_I \bigl(N - N_0\bigr) \,,
\end{align}
 and the solution of Eq.~(\ref{sb:eq-phi-2-slow}) is obtained as
\begin{align}
 {}^{(2)}\phi_I 
 &= \frac{1}{3} D_I^\phi \Bigl[ e^{3 (N - N_0)} - 1 \Bigr]
 + \int_{N_0}^N dN' e^{3 N'} \left( \int^{N'}_{N_0} dN'' e^{- 3 N''}
 \frac{S_I^\phi}{a^2 e^{2 C^\psi} {}^{(0)}V} \right) \,,
\label{sb:sol-phi-2}
\end{align}
 where $N_0$ is an initial time and $C_I^\phi$ and $D_I^\phi$
 represent the initial values of the scalar field and its time derivative. 
Note that the solution ${}^{(0)}\phi_I$ satisfies
the slow-roll conditions (\ref{slow-roll-cond}) 
and (\ref{slow-roll-cond2}) if all the masses are small,
\begin{align}
M^2\equiv \sum_I m_I^2 \ll 1\,.
\label{mass-cond}
\end{align}

Here let us show the time-independence of $S_I^\phi$, which is
given by Eq.~(\ref{sb:sphi}). From Eq.~(\ref{N:Ham-phi}) we have
\begin{align}
 {}^{(0)}K^{2} = 3{}^{(0)}V \,.
\label{sb:sol-K-0}
\end{align}
The leading order potential ${}^{(0)}V$ is given by
\begin{align}
{}^{(0)} V = W \exp \left[ \sum_I m_I {}^{(0)}\phi_I \right]
 = C_V \exp \Bigl[ M^2 \bigl(N - N_0\bigr) \Bigr] \,, 
\end{align}
 where $C_V$ is the initial value of the potential,
\begin{align}
 C_V(x^i) 
\equiv {}^{(0)}V(N_0) = W \exp \left[ \sum_I m_I C_I^\phi \right] \,.  
\label{sb:hatV}
 \end{align}
Substituting the above solution into Eq.~(\ref{sb:sphi}) gives
\begin{align}
 S_I^\phi 
 &= \frac{3 \sqrt{C_V}}{e^{C^\psi}} \pa_i \left(
 \frac{e^{C^\psi}}{\sqrt{C_V}} D^i C^\phi_I \right)
 - \frac{1}{2} \left[ D^2 \bigl(\log C_V\bigr)
 - \frac{1}{2} D^i \bigl(\log C_V\bigr) D_i \bigl(\log C_V\bigr)
 + D^i \bigl(\log C_V\bigr) D^i C^\psi \right] m_I \notag\\
 & \qquad
 - \left[ R - \Bigl(4 D^2 C^\psi + 2D^i C^\psi D_i C^\psi \Bigr)
 + 2 D^i C^\phi_J D_i C^\phi_J \right] m_I \,.
\end{align}
The time-independence of $S_I^\phi$ is now clear
since it is expressed solely in terms of
 $C^\psi$, $C_V$, $C_I^\phi$ and ${}^{(0)}\gamma_{i j}$
which are all time-independent functions.
Therefore we can rewrite the second order solution~(\ref{sb:sol-phi-2})
in a simpler manner as
\begin{align}
 {}^{(2)}\phi_I
 = \frac{1}{3} D_I^\phi \Bigl[ e^{3 (N - N_0)} - 1 \Bigr]
 + \frac{S_I^\phi \, I_\phi (N)}{a^2 e^{2 C^\psi}{}^{(0)}V} \,,
\end{align}
 where
 \begin{align}
 \frac{I_\phi (N)}{ a^2 e^{2 C^\psi} {}^{(0)}V}
 &= \frac{1}{e^{2 C^\psi}} \int_{N_0}^N dN' a^{- 3} \left( \int^{N'}_{N_0} dN''
 \frac{a}{{}^{(0)}V} \right) \,.
\end{align}

Given the solution for the scalar field
up to second order in gradient expansion,
we can now obtain those for $K$, $A_{i j}$ and $\gamma_{i j}$.
At leading order, $K$ is expressed in terms of the scalar field
 through Eq.~(\ref{sb:sol-K-0}) as
 \begin{align}
 {}^{(0)}K = \sqrt{3 {}^{(0)}V}  
 = \sqrt{3 C_V} \exp \left[ \frac{1}{2} M^2 \bigl(N - N_0\bigr) \right] \,.
\label{sol-K-0-2}
 \end{align}
Because of the assumption~(\ref{assum-gamma}),
 ${}^{(0)}\gamma_{i j}$ and ${}^{(0)}A_{i j}$ are trivial,
 \begin{align}
 {}^{(0)}\gamma_{i j}= C_{i j}^\gamma \,, \qquad {}^{(0)}A_{i j} = 0 \,.
 \end{align}
At next-to-leading order, Eq.~(\ref{N:Ham-phi}) becomes
\begin{align}
 {}^{(2)}K &= \sqrt{3{}^{(0)}V} \left\{ \frac{1}{2} \frac{{}^{(2)}V}{{}^{(0)}V}
 + \frac{1}{6} m_I \pa_N {}^{(2)}\phi_I
 - \frac{1}{4 a^2 e^{2 C^\psi} {}^{(0)}V} \Bigl[ R - \Bigl( 4 D^2 C^\psi
 + 2 D^i C^\psi D_i C^\psi \Bigr) + D^i \phi_I D_i \phi^I \Bigr] \right\} \,,
\label{sb:eq-K}
\end{align}
 and Eq.~(\ref{N:eq-Aij}) reduces to
 \begin{align} 
 e^{3 N} \pa_N \Bigl( e^{- 3 N} A_{i j} \Bigr)
 = \frac{1}{a^2 e^{2 C^\psi} {}^{(0)}K} S_{i j}^A \,,
\label{sb:eq-Aij}
 \end{align}
 where $S_{i j}^A$ is independent of time. Specifically,
\begin{align}   
 S_{i j}^A
 &= 3 \Bigl( R_{i j} + D_i C^\psi D_j C^\psi - D_i D_j C^\psi \Bigr)^{T F}
 - \frac{3}{2} \left[ - D_i \Bigl( D_j \bigl(\log C_V\bigr) \Bigr)
 + \frac{1}{2} D_i \bigl(\log C_V\bigr) D_j \bigl(\log C_V\bigr) \right]^{T F}
 \notag\\
 & \qquad - \frac{3}{2} \Bigl[ D_i \bigl(\log C_V\bigr) D_j C^\psi
 + D_j \bigl(\log C_V\bigr) D_i C^\psi + D_i C_I^\phi D_j C_I^\phi \Bigr]^{T F}
 \,. 
 \end{align}

Substituting the solution of the scalar field into Eq.~(\ref{sb:eq-K}),
 we obtain ${}^{(2)}K$ as
\begin{align}
 \frac{{}^{(2)}K}{{}^{(0)}K}
 &= \frac{1}{6} \sum_I m_I D_I^\phi \Bigl[ 2 e^{3 (N - N_0)} - 1 \Bigr]
 + \frac{1}{a^2 e^{2 C^\psi} {}^{(0)}V} S^K \,,
\end{align}
 where 
\begin{align}
 S^K
 &= \frac{1}{2} m_I S_I^\phi I_\phi 
 + \frac{1}{6} m_I S_I^\phi \frac{{}^{(0)}V}{a}
 \int^N_{N_0} dN' \frac{a}{{}^{(0)}V}
 - \frac{1}{4} \Bigl[ R - \Bigl( 4 D^2 C^\psi + 2 D^i C^\psi D_i C^\psi \Bigr)
 + D^i C_I^\phi D_i C_I^\phi \Bigr] \,.
\end{align}
As for $A_{i j}$, we obtain it by integrating Eq.~(\ref{sb:eq-Aij}),
\begin{align}
 A_{i j} 
 &= C_{i j}^A \, e^{3 (N - N_0)}
 + \frac{S^A_{i j}}{a^3 e^{2 C^\psi}} \int_{N_0}^N dN' \frac{a}{{}^{(0)}K} \,.
\end{align}
Finally, ${}^{(2)}\gamma_{i j}$ is given by solving Eq.~(\ref{N:eq-gamma}),
\begin{align}
 {}^{(2)}\gamma_{i j}
 &= - 6 C_{i j}^A \int_{N_0}^N dN' \frac{e^{3 (N' - N_0)}}{{}^{(0)}K}
 - 6 \frac{S_{i j}^A}{e^{2 \psi}} \int_{N_0}^N dN' \frac{1}{a^3 {}^{(0)}K}
 \left( \int_{N_0}^{N'} dN'' \frac{a}{{}^{(0)}K} \right) \,.
\end{align}

Before concluding this subsection, let us fix the remaining gauge degree of
freedom on the uniform $\calN$ slicing mentioned in subsection~\ref{ssec:slice}.
We can make ${}^{(0)}K$ spatially homogeneous on the initial slice by using
this gauge degree of freedom, 
\begin{align}
 C_V(x^i)={}^{(0)}{V}_0 = \ma{const.} \,, 
\label{V0-const}
\end{align}
 where ${}^{(0)}{V}_0$ is a pure constant independent of both space and time.
With this choice, we can substantially simplify the above expressions of
the solution because the spatial derivative of $C_V$ vanishes.

To summarise, thus obtained solution is
\begin{align}
 \phi_I 
 &= C_I^\phi + m_I (N - N_0)
 + \frac{1}{3} D_I^\phi \Bigl[ e^{3 (N - N_0)} - 1 \Bigr]
 + \frac{S_I^\phi I_\phi (N)}{a^2 e^{2 C^\psi}{}^{(0)}V} \,,
 \label{sb:sol-phi}\\
 K &= \sqrt{3 {}^{(0)}V}\left\{ 1
 + \frac{1}{6} m_I D_I^\phi \Bigl[ 2 e^{3 (N - N_0)} - 1 \Bigr]
 + \frac{S^K (N, x^i)}{a^2 e^{2 C^\psi} {}^{(0)}V} \right\} \,,
 \label{sb:sol-K} \\
 \gamma_{i j}
 &= C_{i j}^\gamma - \frac{2 \sqrt{3} C_{i j}^A}{\sqrt{{}^{(0)}V}} e^{3 (N - N_0)}
 I_\gamma (N) - \frac{2 S_{i j}^A J_\gamma (N)}{a^2 e^{2 C^\psi}{}^{(0)}V} \,,
 \label{sb:sol-gamma} \\ 
 A_{i j} 
 &= C_{i j}^A \, e^{3 (N - N_0)}
 + \frac{S^A_{i j} I_A (N)}{\sqrt{3} a^2 e^{2 C^\psi} \sqrt{{}^{(0)}V}} \,,
\label{sb:sol-A}
\end{align}
 where ${}^{(0)}V={}^{(0)}{V}_0 \exp \left[ M^2 (N - N_0) \right]$ and 
$C_I^\phi$, $D_I^\phi$, $C_{i j}^\gamma$ and $C^A_{i j}$
 are initial values of $\phi$, $\pa_N \phi$, $\gamma_{i j}$ and $A_{i j}$,
 respectively. 
The source functions $S_I^\phi$ and $S_{i j}^A$ are time-independent 
whose explicit forms are
\begin{align}
 S_I^\phi 
 &= \frac{3}{e^{C^\psi}} \pa_i \left( e^{C^\psi} D^i C^\phi_I \right)
 - \left[ R - \Bigl(4 D^2 C^\psi + 2D^i C^\psi D_i C^\psi \Bigr)
 + 2 D^i C^\phi_J D_i C^\phi_J \right] m_I \,,
 \label{sb:sol-sphi} \\ 
 S_{i j}^A
 &= 3 \Bigl( R_{i j} + D_i C^\psi D_j C^\psi - D_i D_j C^\psi
 -{1\over 2} D_i C^\phi_ID_j C^\phi_I \Bigr)^{T F} \,,  
 \end{align}
and the functions $I_\phi$, $S^K$, $I_A$, $I_\gamma$ and $J_\gamma$ are 
given by
\begin{align}
 I_\phi (N)
 &= a^2{}^{(0)}V \int_{N_0}^N dN' \frac{1}{a^3}
 \int^{N'}_{N_0} dN'' \frac{a}{{}^{(0)}V} \,, 
\label{Iphi}\\
 S^K (N, x^i)
 &= \frac{1}{2} m_I S_I^\phi I_\phi (N)
 + \frac{1}{6} m_I S_I^\phi \frac{{}^{(0)}V}{a} \int^N_{N_0} dN' 
\frac{a}{{}^{(0)}V}
 - \frac{1}{4} \Bigl[ R - \Bigl( 4 D^2 C^\psi + 2 D^i C^\psi D_i C^\psi \Bigr)
 + D^i C^\phi_I D_i C^\phi_I \Bigr] \,, 
\label{SKdef}\\
 I_\gamma (N)
 &= a^3 \sqrt{{}^{(0)}V} \int_{N_0}^N dN' \frac{1}{a^3 \sqrt{{}^{(0)}V}} \,, 
\label{Igamma}\\
 J_\gamma (N)
 &= a^2 {}^{(0)}V \int_{N_0}^N dN' \frac{1}{a^3 \sqrt{{}^{(0)}V}}
 \left( \int_{N_0}^{N'} dN'' \frac{a}{\sqrt{{}^{(0)}V}} \right) \,, 
\label{Jgamma}\\
 I_A (N)
 &= \frac{\sqrt{{}^{(0)}V}}{a} \int_{N_0}^N dN' \frac{a}{\sqrt{{}^{(0)}V}} \,.
\label{IAdef}
\end{align}

%%%%%%%%%%%%%%%%%%%%%%%%%%%%%%%%%%%%%%%%%%%%%%%%%%%%%%%%%%%%%%%%%%%%%%%%%%
\subsection{Solution on the uniform $K$ slice}
\label{subsec:uniformK}

In this subsection, we derive the solution on the $K$ gauge 
 by applying a gauge transformation to the solution on the $\calN$ gauge. 
To do so, we first need to determine the generator of the gauge transformation
 between the two slices, $N \to \tilde{N} = N + n (N, x^i)$ or conversely
 $\tilde{N} + \tilde{n} (\tilde{N}, \tilde{x}^i) = N$. The nonlinear
 gauge transformation is discussed in detail in Appendix~\ref{app:gauge}. 

As is clear from Eq.~(\ref{sol-K-0-2}) with the initial condition
$C_V={}^{(0)}V_0=$const., Eq.~(\ref{V0-const}), 
the uniformity of $K$ on the uniform $\calN$ slices is kept in this model
 at leading order if it is chosen so on the initial slice.
Thus no gauge transformation is necessary at leading order.

At next-to-leading order, the gauge transformation of $K$ is from 
Eq.~(\ref{GT-K-2}) given by
\begin{align}
 {}^{(2)}\tilde{K} 
 &= {}^{(2)}K + {}^{(2)}\tilde{n} \pa_N {}^{(0)}K \,.
\end{align}
Since ${}^{(2)}\tilde{K}$ vanishes on the uniform $K$ slice,
 ${}^{(2)}\tilde{n}$ is determined from Eq.~(\ref{sb:eq-K}) as
\begin{align}
{}^{(2)}\tilde{n}
 &= - \frac{2}{M^2} \left\{ \frac{1}{6} m_I D_I^\phi \Bigl[ 2 e^{3 (N - N_0)}
 - 1 \Bigr] + \frac{1}{a^2 e^{2 C^\psi} {}^{(0)}V} S^K \right\} \,.
\end{align}

The solution for $\psi$ is obtained from Eq.~(\ref{GT-psi-2}) as
\begin{align}  
 \tilde{\psi}
 &= \psi - {}^{(2)}\tilde{n} + \calO (\epsilon^4) \notag\\
 &= C^\psi + \frac{2}{M^2} \left\{ \frac{1}{6} m_I D_I^\phi \Bigl[
 2 e^{3 (N - N_0)} - 1 \Bigr] + \frac{1}{a^2 e^{2 C^\psi}{}^{(0)}V} S^K \right\} \,,
\label{sol-psi-onK}
\end{align}
where $C^\psi$ is given in (\ref{N:sol-psi}) 
and that for $\gamma_{i j}$ is same as $\gamma_{i j}$ on the $\calN$
 gauge from Eq.~(\ref{GT-gamma-2}),
\begin{align}  
 \tilde{\gamma}_{i j} 
 &= \gamma_{i j} + \calO (\epsilon^4) \,.  
\end{align}

%%%%%%%%%%%%%%%%%%%%%%%%%%%%%%%%%%%%%%%%%%%%%%%%%%%%%%%%%%%%%%%%%%%%%%%%%%
\subsection{The curvature perturbation}

At next-to-leading order in gradient expansion,
 we have to extract the scalar component $\chi$ from $\gamma_{ij}$ 
to obtain the curvature perturbation $\frR$. 
In linear theory, the variable $\chi$ reduces to the traceless
scalar-type component $H_T^{\rm Lin}$, and the curvature perturbation is
given by ${\cal R}^{\rm Lin}=(H^{\rm Lin}_L+{H^{\rm Lin}_T/3})Y$, 
where $\psi\to H_L^{\rm Lin}$ in the linear limit, and
we have followed the notation of \cite{Kodama:1985bj}.

Neglecting the contribution of gravitational waves and
 focusing on that of the scalar-type perturbations,
 we can assume that ${}^{(0)}\gamma_{i j}$ in the $K$ gauge 
 approaches the flat metric at late times when the adiabatic limit is reached,
\begin{align}
 {}^{(0)}\gamma_{i j} ~ \to ~ \delta_{i j} \quad \bigl(N \to 0\bigr) \,.
\label{cond:gamma-flat}
\end{align}
This condition completely fixes the remaining spatial gauge degrees
of freedom, say, within the class of the time-slice-orthogonal threading.  

Under this condition, we rewrite $\gamma_{i j}$ on the $K$ gauge 
given in Eq.~(\ref{sb:sol-gamma}) to manifest its time and spatial dependence,
\begin{align}  
 \gamma_{i j}
 &= \delta_{i j} + 2 \sqrt{3} C_{i j}^A (x^i) \, f^A (0)
 + C_{i j}^\chi (x^i) \, f^\chi (0) - 2 \sqrt{3} C_{i j}^A (x^i) \, f^A (N)
 - C_{i j}^\chi (x^i) \, f^\chi (N) \,,  
\label{cond:gamma}
\end{align}
 where $C^\gamma_{ij}=\delta_{ij}$ from Eq.~(\ref{cond:gamma-flat}),
and $C_{i j}^\chi$ is given by 
 \begin{align}
 C_{i j}^\chi
 = 6 e^{- 2 C^\psi} \left( \pa_i C^\psi \pa_j C^\psi - \pa_i \pa_j C^\psi
 - \frac{1}{2} \pa_i C_I^\phi \pa_j C_I^\phi \right)^{T F} \,.
\end{align}
The time-dependent functions $f^A$ and $f^\chi$ are given by
\begin{align}
 f^A (N) = \frac{e^{3 (N - N_0)}}{\sqrt{{}^{(0)}V}} I_\gamma (N) \,, \qquad
 f^\chi (N) &= \frac{1}{a^2\,{}^{(0)}V} J_\gamma (N) \,,
\end{align}
where the functions $I_\gamma$ and $J_\gamma$ are defined in Eqs.~(\ref{Igamma})
and (\ref{Jgamma}), respectively.

Now we extract the scalar part from $C_{i j}^{A}$ and  $C_{i j}^{\chi}$
by using the definition of $\chi$ given by Eq.~(\ref{sb:def-chi}).
We note that this definition is unique in the sense that it reduces to the
standard scalar part in the limit of linear theory and gives the
$\calO(\epsilon^2)$ correction unambiguously.

The contribution of $C_{i j}^{A}$ to $\chi$ may be determined by evaluating 
Eq.~(\ref{N:Mom}) on the initial slice with the help of
Eqs.~(\ref{sb:sol-phi}) and (\ref{sb:sol-A}),
\begin{align}
 e^{- 3 C^\psi} \pa^j \Bigl( e^{3 C^\psi} C^A_{i j} \Bigr)
 = \frac{2}{3} \pa_i {}^{(2)}K (N_0)
 - \sqrt{\frac{{}^{(0)}V_0}{3}} D_I^\phi \pa_i C_I^\phi \,,
\label{sb:eq-CAij}
\end{align}
 where
\begin{align}  
 {}^{(2)}K (N_0) &= \sqrt{3 {}^{(0)}{V}_0} \left( \frac{1}{6} m_I D_I^\phi
 + \frac{ 4 \triangle C^\psi + 2 \pa^i C^\psi \pa_i C^\psi
 - \pa^i C^\phi_I \pa_i C^\phi_I}{4 a^2 (N_0) e^{2 C^\psi} {}^{(0)}{V}_0} 
\right) \,.
\end{align}
Applying the definition of $\chi$ in Eq.~(\ref{sb:def-chi}) to the above,
we find the contribution from $C_{i j}^{A}$ is 
\begin{eqnarray}
\chi^A=- \sqrt{3} {}^{(2)}K (N_0) \Bigl( f^A (0) - f^A (N) \Bigr)
 + \frac{3}{2} \sqrt{{}^{(0)}V_0} \triangle^{-1} \pa^i \Bigl( D_I^\phi \pa_i C_I^\phi
 \Bigr) \Bigl( f^A (0) - f^A (N) \Bigr)\,.
\label{chiA}
\end{eqnarray}
As for the contribution of $C_{i j}^{\chi}$ to $\chi$,
we can make use of the relation,
\begin{align}  
 \frac{1}{6} \triangle R \Bigl[ e^{2 C^\psi} \delta_{i j} \Bigr]
 = \pa^i \Bigl\{ e^{- 3 C^\psi} \pa^j \Bigl[ e^{C^\psi} \Bigl(
 \pa_i C^\psi \pa_j C^\psi - \pa_i \pa_j C^\psi \Bigr)^{T F} \Bigr] \Bigr\} \,,
\label{sb:eq-Cgij}
\end{align}
where the left-hand side is the Ricci scalar of $e^{2 C^\psi} \delta_{i j}$. 
Thus the contribution from $C_{i j}^{\chi}$ is
\begin{eqnarray}
\chi^\chi =- \frac{3}{4} R \Bigl[ e^{2 C^\psi} \delta_{i j} \Bigr]
 \Bigl( f^\chi (0) - f^\chi (N) \Bigr) 
 + \frac{9}{4} \triangle^{-1} \Bigl\{ \pa^i e^{- 3 \psi} \pa^j
 \Bigl[ e^{C^\psi} \Bigl( \pa_i C_I^\phi \pa_j C_I^\phi \Bigr)^{T F} \Bigr]
 \Bigr\} \Bigl( f^\chi (0) - f^\chi (N) \Bigr) \,.  
\label{chichi}
\end{eqnarray}
Adding both contributions together, we obtain
\begin{align}
\chi =\chi^A+\chi^\chi\,,
\label{sol-E-onK}
\end{align}
where $\chi^A$ and $\chi^\chi$ are given, respectively, by
Eqs.~(\ref{chiA}) and (\ref{chichi}).

Finally summing all the contributions to the curvature perturbation 
$\frR=\psi_K+{\chi_K/ 3}$, where $\psi_K$ and $\chi_K$ are given by 
Eqs.~(\ref{sol-psi-onK}) and (\ref{sol-E-onK}), respectively, we
obtain 
\begin{align}
 \frR_K
 &= C^\psi + \frac{2}{M^2} \left\{ \frac{1}{6} m_I D_I^\phi \Bigl[
 2 e^{3 (N - N_0)} - 1 \Bigr] + \frac{1}{a^2 e^{2 C^\psi} {}^{(0)}V} S^K \right\}
- \frac{1}{\sqrt{3}} {}^{(2)}K (N_0) \Bigl( f^A (0) - f^A (N) \Bigr) \notag\\
 & \qquad 
 - \frac{1}{4} R \Bigl[ e^{2 C^\psi} \delta_{i j} \Bigr]
 \Bigl( f^\chi (0) - f^\chi (N) \Bigr) 
 + \frac{\sqrt{{}^{(0)}V_0}}{2} \triangle^{-1} \pa^i \Bigl( D_I^\phi \pa_i C_I^\phi
 \Bigr) \Bigl( f^A (0) - f^A (N) \Bigr) \notag\\
 & \qquad 
 + \frac{3}{4} \triangle^{-1} \Bigl\{ \pa^i e^{- 3 \psi} \pa^j \Bigl[
 e^{C^\psi} \Bigl( \pa_i C_I^\phi \pa_j C_I^\phi \Bigr)^{T F} \Bigr] \Bigr\}
 \Bigl( f^\chi (0) - f^\chi (N) \Bigr) \,.  
\end{align}

What we need to know is the final value of $\frR$ at sufficiently late times, 
$N\to 0\ (a\to a_0 e^{N_0})$. We take $N_0$ to be a time around which the
scales relevant to cosmological observations crossed the Hubble horizon, 
hence $N_0\gtrsim50$.
In this case, at $N=0$, the curvature perturbation reduces to
\begin{align}
\frR_K(N=0)\approx {}^{(0)}C^{\psi}+{}^{(2)}C^{\psi}-{m_I\over 3 M^2}D^\phi_I
\,.\label{sb:late-time-R}
\end{align}
The first term, ${}^{(0)}C^{\psi}$ represents the leading order
curvature perturbation obtainable in the usual $\delta N$ formalism,
and the remaining terms represent $\calO(\epsilon^2)$ contributions, 
the calculation of which is the main purpose of the beyond $\delta N$ formalism.
As a check, we have confirmed that the above result is consistent with the
one obtained in linear perturbation theory with the same background solution.

It should be noted that there is no contribution from $\chi_K$ at $N=0$ by
definition. However, this does not mean the evaluation of $\chi_K$ was
meaningless. In general it can make an important contribution to $\frR$
around the horizon crossing time $N\sim N_0$, and by matching our $\frR$ 
with that evolved from inside the horizon at $N\sim N_0$, the final 
value $\frR(N=0)$ is determined both by the values of $\psi_K$ and
$\chi_K$ and by their derivatives at $N\sim N_0$.

It may be noted that the last term proportional to $D^\phi_I$
comes the time derivative of $\phi_I$ at $N=N_0$, 
$\pa_N\phi_I(N_0)=D^\phi_I=\calO(\epsilon^2)$.
Although this is generically small by construction, the contribution 
of the term itself can become large since it is proportional to 
$m_I/M^2\sim 1/M$ where $M^2\ll 1$ as assumed in Eq.~(\ref{mass-cond}).

%%%%%%%%%%%%%%%%%%%%%%%%%%%%%%%%%%%%%%%%%%%%%%%%%%%%%%%%%%%%%%%%%%%%%%%%%%
\section{Summary and discussion}
\label{sec:summary}

In this paper, we developed a theory of nonlinear cosmological
 perturbations on superhorizon scales in the context of inflationary 
cosmology. We considered a multi-component scalar field
 with a general kinetic term and a general form of the potential.
To discuss the superhorizon dynamics, we employed the ADM formalism
 and the spatial gradient expansion approach.

Different from the single-field case, there is a difficulty in 
solving the equations in the multi-field case.
At leading-order, the equations take the same form as
those for the homogeneous and isotropic FLRW background
with suitable identifications of variables.
In particular, there are correspondences between the cosmic time
in the background and the proper time along each comoving trajectory,
$\tau \Leftrightarrow t$, and the scale factor with the one defined
locally $a e^\psi \Leftrightarrow a$.
This implies that given the background solution, 
the solution at leading-order in gradient expansion is automatically
obtained.

In the single-field case, one can show the coincidence 
among the comoving, uniform expansion, and synchronous slicings 
at leading order. This allows us to set the lapse function to unity,
and replace the proper time by the cosmic time
 in the solution for the scalar field. 
Then the metric is expressed in terms of the scalar field easily,
and the next-to-leading order equations can be solved straightforwardly
because the space-time dependence of the source terms consisting of 
the leading order quantities is explicitly known.

In cosmological perturbation theory, the most important quantity
to be evaluated is the curvature perturbation on the comoving
 slices which is conserved on superhorizon scales after the universe 
has reached the adiabatic limit. This quantity accurate to next-to-leading
order may be relatively easily obtained in the single-field case because
of the above mentioned coincidence among the comoving, uniform expansion
and synchronous slicings.

On the other hand, in the multi-field case, such a coincidence 
between different slicings does not hold.
This implies the following. One can express the lapse function
 as a function of the scalar field, but the scalar field
is also a function of the proper time. Thus one has the equation, 
\begin{gather}
 \alpha = f \Bigl[ t, \, \phi (\tau) \Bigr]
 = f \left[ t, \, \phi \left( \int \alpha \, dt \right) \right]\,.
\end{gather}
To go beyond the leading order, one needs to solve this equation 
for $\alpha$, but it seems almost impossible.

In this paper, we developed a formalism to go beyond the leading order
which avoids the above problem.
Namely, we first solve the field equations in a slicing in which
the lapse function is trivial. The synchronous slicing is one of
such slicings, but we adopt the uniform $e$-folding number slicing
in which the time slices are chosen in such a way that
the number of $e$-folds along each orbit orthogonal to the time
slices, $\calN$, is spatially homogeneous on each time slice.
In other words, the uniform $\calN$ slicing is a synchronous
 slicing if $\calN$ is used as the time coordinate.

In this slicing we can solve the equations to next-to-leading order
without encountering the above mentioned problem.
After the solution to next-to-leading order is obtained, we
transform it to the one in the uniform expansion slicing
which is known to be identical to the comoving slicing on
superhorizon scales in the adiabatic limit.
Thus the gauge transformation laws play an essential role
in our formalism. We derived them which are accurate to
next-to-leading order.
Note that they are fully nonlinear in nature in the language 
of the standard perturbation approach. They are summarised in 
Appendix~\ref{app:gauge}.

 As a demonstration of our formalism, we considered an analytically 
solvable model and constructed the explicit form of the solution. 
Namely, we considered a multi-component canonical scalar field with
exponential potential, $V(\phi_I)=W\exp[\sum_J m_J\phi_J]$.
Following the general procedure discussed above, we first solved for 
the scalar field and the metric in the uniform $\calN$ slicing. 
Then using a remaining gauge degree of freedom of this slicing, we 
set the initial condition such that it coincides with the uniform expansion
slicing at leading order. In this slicing with this initial condition,
the next-to-leading order solution was straightforwardly found.
Then we applied the derived nonlinear gauge transformation to it
to obtain the solution in the uniform expansion slicing.
Finally, from thus obtained spatial metric which takes the form,
$e^\psi\gamma_{ij}$ where $\gamma_{ij}$ is a unimodular metric,
we constructed the generalised, nonlinear curvature perturbation
$\frR$ defined by
\begin{align}
\frR=\psi+{\chi\over 3}\,; \quad
\chi \equiv - \frac{3} {4}  \triangle^{-1}  
\Bigl\{\pa^i e^{-3\psi} \pa^j \Bigl[e^{3\psi} \bigl(\gamma_{ij}-\delta_{ij}\bigr)
 \Bigr] \Bigr\}\,.
\label{summary-frR}
\end{align}

By inspecting the form of the final value of the curvature perturbation,
we argued that the decaying modes of the scalar field may give rise to
non-negligible contribution to the final, conserved curvature perturbation.
To quantify the effect, it is necessary to match our solution on super-horizon
scales with the one solved from sub-horizon scales. The evaluation
of these next-to-leading order corrections to the spectrum
as well as to the bispectrum of the curvature perturbation is left for
future study.

%%%%%%%%%%%%%%%%%%%%%%%%%%%%%%%%%%%%%%%%%%%%%%%%%%%%%%%%%%%%%%%%%%%%%
\acknowledgments
AN is grateful to Shinji Mukohyama and Yuki Watanabe
 for valuable comments and fruitful discussions.
YT would like to thank Ryo Saito, Teruaki Suyama, 
Jun'ichi Yokoyama, Kiyoung Choi, 
Seoktae Koh and Shuichiro Yokoyama for their comments 
and discussions on this work. 
This work is supported in part by Monbukagaku-sho 
Grant-in-Aid for the Global COE programs, 
``The Next Generation of Physics, Spun from Universality 
and Emergence'' at Kyoto University.
AN is partly supported by Grant-in-Aid for JSPS Fellows No.~21-1899 and
 JSPS Postdoctoral Fellowships for Research Abroad. 
YT also wishes to acknowledge financial support
 by the RESCEU, University of Tokyo and 
by JSPS Grant-in-Aid for Young Scientists
(B) No.~23740170 and for JSPS Postdocoral Fellowships. 
The work of MS is supported by JSPS Grant-in-Aid for
 Scientific Research (A) No.~21244033,
and by Grant-in-Aid for Creative Scientific Research No.~19GS0219.

%%%%%%%%%%%%%%%%%%%%%%%%%%%%%%%%%%%%%%%%%%%%%%%%%%%%%%%%%%%%%%%%%%%%%%%%%%
\appendix

\section{Coincidence of four slices in single scalar case}
\label{app:coincide}

In the single-field case, it can be shown that the four slicings, 
the comoving, uniform $K$, synchronous and uniform $\calN$ slicings,
coincide with each other at leading order in gradient expansion.
Here we demonstrate it. For simplicity, we consider a canonical scalar field,
 but the generalisation is straightforward.
 
At leading order in gradient expansion,
the Hamiltonian and momentum constraints are written as
\begin{align}
 \frac{1}{3} K^2 &= \frac{1}{2} \bigl(\pa_\tau \phi\bigr)^2 + V (\phi) \,,
\label{appc:Ham} \\
 \pa_i K &= - \frac{3}{2} \pa_\tau \phi \pa_i \phi \,,
\label{appc:Mom} 
\end{align}
 where $\tau$ is defined in terms of the lapse function as 
 \begin{align}
 \tau (t, x^i) \equiv \int_{x^i = const.} \alpha (t', x^i) \, dt' \,. 
 \end{align}
Here let us choose the uniform $K$ slicing, 
\begin{gather}
 K (t, x^i) = K (t) \,.
\end{gather}
Then Eq.~(\ref{appc:Mom}) immediately implies that
 the scalar field is uniform on this slice,
\begin{gather}
 \phi (t, x^i) \quad \to \quad \phi = \phi (t) \,.  
\end{gather}
This shows the uniform $K$ slice coincides with the comoving slice. 

Now since the left-hand side of Eq.~(\ref{appc:Ham}) as well as
the potential term on the right-hand side is homogeneous, it follows
that the kinetic term is homogeneous. This means
\begin{gather}
 \pa_\tau \phi = uniform \quad \to \quad \alpha = \alpha (t) \,.
\end{gather}
Therefore the uniform $K$ slicing coincides with the synchronous slicing.
Then the spatial homogeneity of both $K$ and $\alpha$ implies that of
the number of $e$-folds $\calN$ by definition, Eq.~(\ref{def-calN}).
Thus we have shown the coincidence of these four slicings
at leading order in gradient expansion. 

As we noticed in subsection~\ref{ssec:slice}, there is
 a remaining gauge degree of freedom in the synchronous and uniform $\calN$
slicings. In the above proof, this freedom is tacitly fixed 
since we took the uniform $K$ slicing as a starting point.
In other words, the uniform $K$ (or comoving) slicing implies
it is synchronous and uniform $\calN$. However, if one starts
from the synchronous or uniform $\calN$ slicing, we have to set 
the initial data such that $K$ or $\phi$ is uniform on the initial slice
in order to show the coincidence.
In fact, if we use this remaining degree of freedom 
to set a different condition on the initial data,
for example $\psi = 0$, the synchronous or uniform $\calN$ slicing
will not coincide with the uniform $K$ or comoving slicing.

In passing let us also comment on the conservation of the curvature 
perturbation. From the definition of $K$, Eq.~(\ref{def-K}), one has
\begin{align}
 \pa_t \psi (t, x^i) = - H (t) + \frac{1}{3} \alpha (t, x^i) K (t, x^i) \,.
\label{app:eq-cons}
\end{align}
This shows how the inhomogeneous part on the right-hand side
 contributes to the evolution of the curvature perturbation.
As we have discussed above, there is no such inhomogeneities
in the uniform $K$ or comoving slicing in the single-scalar case. 
This is a proof of the conservation of the nonlinear
curvature perturbation on super-horizon scales.

On the other hand, in the case of multi-field,
 the corresponding scalar component on the right-hand side
of the momentum constraint, Eq.~(\ref{appc:Mom}), becomes uniform on the 
uniform $K$ slice. However this scalar component does not appear 
in Eq.~(\ref{appc:Ham}) in general, and hence one cannot show 
the homogeneity of the lapse function as in the single-field case.

We note that this does not mean that the Hamiltonian and momentum
constraints are not related. Actually the leading order
momentum constraint reduces to the spatial derivative of
 the Hamiltonian constraint in the slow-roll limit
 as shown in Appendix~\ref{sapp:slow}.
For a non-slow roll case, this relation between the Hamiltonian
and momentum constraints holds only if we take into account the
next-to-leading order corrections properly (see Appendix~\ref{sapp:nonslow}).

%%%%%%%%%%%%%%%%%%%%%%%%%%%%%%%%%%%%%%%%%%%%%%%%%%%%%%%%%%%%%%%%%%%%%%%%%%
\section{Uniform $K$ slicing}
\label{app:uniK}

In this Appendix, we derive the basic equations
 on the uniform $K$ slicing, $K (t, x^i) = 3 H (t)$,
 and discuss a difficulty in constructing the solution at next-to-leading 
 order in gradient expansion. 

In the uniform $K$ slicing and time-slice-orthogonal threading, 
 the constraint equations are given by
 \begin{align}
 \frac{1}{a^2 e^{2 \psi}} \Bigl[ R - \Bigl( 4 D^2 \psi + 2D^i \psi D_i \psi 
 \Bigr) \Bigr] + 6 H^2
 &= 2 E \,,
\label{K:Ham}  \\
 e^{- 3 \psi} D_j \Bigl( e^{3 \psi} A^j{}_i \Bigr)
 &=  2 P_{(I J)} \pa_\tau \phi^I \pa_i\phi^J \,.
 \end{align}
The evolution equation for $K$ and $A_{i j}$ are
\begin{align}
 3 \pa_\tau H
 &= - \frac{3}{2} (E + P)
 - \frac{1}{2 a^2 e^{2 \psi}} \Bigl[ R - \Bigl( 4 D^2 \psi
 + 2 D^i \psi D_i \psi \Bigr) \Bigr] \notag\\
 & \qquad
 + \frac{1}{a^2 e^{2\psi} \alpha} \Bigl( D^2 \alpha
 + D^i \alpha D_i \psi - \alpha P _{(I J)} D^i \phi^I D_i \phi^J \Bigr)
\label{K:eq-K}\,, \\
 \pa_\tau A_{i j}
 &= - 3 H A_{i j}
 - \frac{1}{a^2 e^{2\psi}} \Bigl( R_{i j} + D_i \psi D_j \psi
 -  D_i D_j \psi \Bigr)^{T F} \notag\\
 & \qquad
 + \frac{1}{a^2 e^{2\psi} \alpha} \Bigl( D_i D_j \alpha
 - D_i \alpha D_j \psi - D_j \alpha D_i \psi
 - 2 \alpha P_{(I J)} D_i \phi^I D_j \phi^J \Bigr)^{TF} \,,
\end{align}
 where
 \begin{align}
 E = 2 P_{(I J)} \pa_\tau \phi^I \pa_\tau \phi^J - P \,, 
 \end{align}
 and $\tau$ is defined in terms of the lapse function as 
 \begin{align}
 \tau (t, x^i) \equiv \int_{x^i = const.} \alpha (t', x^i) \, dt' \,, 
 \end{align}
and the equation for the scalar field is
\begin{gather}
 \pa_\tau \Bigl( P_{(I J)} \pa_\tau \phi^J \Bigr) + 3 H P_{(I J)} \pa_\tau \phi^J 
 - \frac{1}{a^2\alpha e^{3\psi}} \pa_i \Bigl( \alpha e^\psi P_{(I J)} D^i \phi^J
 \Bigr) - \frac{1}{2} P_{I} = 0 \,.
\label{K:eq-scalar}
\end{gather} 

First, let us consider the leading order.
The Hamiltonian constraint~(\ref{K:Ham}) reduces to 
\begin{align}
 3 H^2 = {}^{(0)}E + \calO (\epsilon^2) \,,
\end{align}
hence ${}^{(0)}E$ is homogeneous on this slice.
This means the solution for ${}^{(0)}E$ exactly coincides with 
 the background solution $E_{BG}$.
As for $P$, the leading order solution is easily constructed in terms of
 the background solution once it is known as 
discussed in Sec.~\ref{ssec:leading}.
To be specific, if the background pressure is expressed as a function of $t$, 
\begin{align}
 P_{BG} (t) = P_{BG} \Bigl[ t, P_0 (t_0) \Bigr] \,,
\end{align}
 the leading order solution for $P$ is given by
 \begin{align}
{}^{(0)}P(t, x^i) = P_{BG} \Bigl[ \tau, P_0 (\tau_0) \Bigr] \,.   
\label{K:sol-P}
 \end{align}

The lapse function can be obtained from Eq.~(\ref{K:eq-K}) as
\begin{gather}
 {}^{(0)}\alpha = - 2 \dfrac{dH (t)}{dt}\frac{1}{{}^{(0)}E(t) +{}^{(0)}P(t, x^i)}
 + \calO (\epsilon^2) \,.
\label{K:sol-alpha}
\end{gather}
Thus the lapse function becomes inhomogeneous if ${}^{(0)}P$ is so.
From Eq.~(\ref{def-K}), the evolution of $\psi$ is related with 
the inhomogeneity of ${}^{(0)}\alpha$ in this slicing as
 \begin{align}
 \pa_t \psi = H (\alpha - 1) \,,
\label{K:eq-psi}
 \end{align}
 which reflects the well-known fact that the curvature perturbation
 evolves if a non-adiabatic pressure exists.
However there is a profound problem in Eq.~(\ref{K:sol-alpha}).

Although it looks like Eq.~(\ref{K:sol-alpha}) gives the
solution for ${}^{(0)}\alpha$, it is not so because
the right-hand side depends also on ${}^{(0)}\alpha$.
In reality, Eq.~(\ref{K:sol-P}) gives ${}^{(0)}P$ as a function of 
$\tau$, which is given by the time integration of the lapse function.
Therefore Eq (\ref{K:sol-alpha}) can be schematically expressed as 
\begin{gather}
 \alpha = f \Bigl[ t, P (\tau) \Bigr]
 = f \left[ t, P \left( \int \alpha \, dt \right) \right] \,.
\end{gather}
In general it is very difficult to solve this equation, at least
analytically.

As clear from Eq.~(\ref{K:eq-K}),
 one needs to evaluate the spatial derivatives of $\alpha$ for example 
 at next-to-leading order. But this will be almost impossible.
This is the main problem in constructing a solution at next-to-leading order
 in gradient expansion on the uniform $K$ slicing. 
This kind of problem arises also on other slicings
 except for the synchronous or uniform $\calN$ slicings. 
This is why the uniform $\calN$ slicing is used in our formalism.

As a closing remark of this Appendix, 
 we emphasise that the above difficulty becomes actually problematic
 only at next-to-leading order. It is not a problem at leading order.
This is because the solution of the ``curvature perturbation''
 ${}^{(0)}\psi$ is given by a time integration which depends
only on the initial and final values of $\calN$. From Eq.~(\ref{K:eq-psi})
we have 
\begin{gather}
 {}^{(0)}\psi(t, x^i) - {}^{(0)}\psi(t_0, x^i)
 = \int_{t_0}^t dt' \, \bigl( {}^{(0)}\alpha - 1 \bigr) H
 = \calN(t,x^i)-\calN(t_0,x^i)-N(t)+N(t_0)\,.
\end{gather}
Thus, at leading order in gradient expansion,
as far as we are interested in the curvature perturbation
there is no need to know the solution for ${}^{(0)}\alpha$.
This is why the $\delta N$ formalism works well
despite the presence of the above problem.

%%%%%%%%%%%%%%%%%%%%%%%%%%%%%%%%%%%%%%%%%%%%%%%%%%%%%%%%%%%%%%%%%%%%%%%%%%
\section{Nonlinear gauge transformation}
\label{app:gauge}

We derive the gauge transformation rules for the metric,
 its derivative ($K$ and $A_{i j}$) and the scalar field.
We consider a nonlinear gauge transformation from a coordinate system 
 with vanishing shift vector $\beta^i = 0$,
 to another coordinate system in which the new shift vector also vanishes,
 $\tilde{\beta}^i = 0$. We note that once the time slicing is changed,
 the shift vector appears in the new slicing in general.
So the spatial coordinates also need to be changed to 
 eliminate thus appeared shift vector.

We use the background $e$-folding number $N$ as the time coordinate and
 define the temporal and spatial shift, $n$ and $L^i$, respectively, 
\begin{align}
 N &\to \tilde{N} = N + n (N, x^i) \,, \notag\\
 x^i &\to \tilde{x}^i = x^i + L^i (N, x^i) \,, \notag
 \end{align}
 or conversely
\begin{align}
 \tilde{N} + \tilde{n} (\tilde{N}, \tilde{x}^i) = N \,,
 \label{GT:n} \\
 \tilde{x}^i + \tilde{L}^i (\tilde{N}, \tilde{x}^i) = x^i \,.
 \end{align}
Under the change of the coordinates, the line element should remain invariant,
\begin{align}
ds^2
 &= - \frac{\alpha^2}{H^2 (N)} d N^2
 + a^2 (N) e^{2 \psi} \gamma_{i j} dx^i dx^j
 = - \frac{\tilde{\alpha}^2}{H^2 (\tilde{N})} d \tilde{N}^2
 + a^2 (\tilde{N}) e^{2 \tilde{\psi}} \tilde{\gamma}_{i j}
 d \tilde{x}^i d \tilde{x}^j \,.
\end{align}
Equating the coefficients of $d \tilde{N}^2$,
$d \tilde{N}d \tilde{x}^i$, and $d \tilde{x}^i d \tilde{x}^j$
on both sides of the above, we obtain
\begin{align}
 \frac{\tilde{\alpha}^2}{H^2 (\tilde{N})}
 &= \frac{\alpha^2}{H^2 (N)} \bigl(1 + \pa_{\tilde{N}} \tilde{n}\bigr)^2
 - a^2 (N) e^{2 \psi} \gamma_{i j} \pa_{\tilde{N}} \tilde{L}^i
 \pa_{\tilde{N}} \tilde{L}^j \,,
\label{GT-alpha} \\
 0 &= \frac{\alpha^2}{H^2 (N)} \bigl(1 + \pa_{\tilde{N}} \tilde{n}\bigr)
 \pa_{\tilde{i}} \tilde{n} 
 - a^2 (N) e^{2 \psi} \gamma_{k j} \pa_{\tilde{N}} \tilde{L}^j
 \bigl( \delta^k{}_i + \pa_{\tilde{i}} \tilde{L}^k \bigr) \,,
\label{GT-beta} \\
 a^2 (\tilde{N}) e^{2 \tilde{\psi}} \tilde{\gamma}_{i j} 
 &= a^2 (N) e^{2 \psi} \gamma_{k l}
 \bigl( \delta^k{}_i + \pa_{\tilde{i}} \tilde{L}^k \bigr)
 \bigl( \delta^l{}_j + \pa_{\tilde{j}} \tilde{L}^l \bigr)
 - \frac{\alpha^2}{H^2 (N)} \pa_{\tilde{i}} \tilde{n} \pa_{\tilde{j}} \tilde{n} \,.
\label{GT-gamma} 
\end{align}

\subsection{Leading order gauge transformation}
First we derive the leading order gauge transformation.
To begin with, we note the spatial shift $\tilde{L}^i$ is 
$\calO (\epsilon)$ from Eq.~(\ref{GT-beta}),
 since it is proportional to the spatial derivative of $\tilde{n}$,
 \begin{gather}
 \pa_{\tilde{N}} \tilde{L}^i \sim \gamma^{i j} \pa_{\tilde{i}} \tilde{n}
 = \calO (\epsilon) \,. 
\label{order-shift}
 \end{gather}
Hence neglecting $\pa_{\tilde{N}} \tilde{L}^i$ in Eq.~(\ref{GT-alpha}),
we have
\begin{gather}
 \frac{\alpha^2}{H^2 (N)} \bigl( 1 + \pa_{\tilde{N}} \tilde{n} \bigr)^2
 = \frac{\tilde{\alpha}^2}{H^2 (\tilde{N})} + \calO (\epsilon^2) \,.
\end{gather}
This gives the leading order gauge transformation of the lapse function,
 \begin{align}
 \tilde{\alpha} (\tilde{N}, \tilde{x}^i)
 &= \frac{H (\tilde{N})}{H (N)} \bigl( 1 + \pa_{\tilde{N}} \tilde{n} \bigr)
 ~ \alpha (N, x^i) + \calO (\epsilon^2) \notag\\
 &= \frac{H (\tilde{N})}{H (\tilde{N} + \tilde{n})}
 \bigl(1 + \pa_{\tilde{N}} \tilde{n} \bigr)
 ~ \alpha (\tilde{N} + \tilde{n}, \tilde{x}^i)
 + \calO (\epsilon^2) \,.
 \end{align} 

The gauge transformation rules for the spatial metric components,
$\psi$ and $\gamma_{i j}$, are derived from Eq.~(\ref{GT-gamma}).
Taking the determinant of Eq.~(\ref{GT-gamma})
 and neglecting $\pa_{\tilde{N}} \tilde{L}^i$ and $\pa_{\tilde{i}} \tilde{n}$,
we obtain
\begin{gather}
 a^2 e^{2 \psi} = a^2 (\tilde{N}) e^{2 \tilde{\psi}} + \calO (\epsilon^2) \,.
\end{gather}
This gives the leading order gauge transformation of $\psi$ as
 \begin{align}
 \tilde{\psi} (\tilde{N}, \tilde{x}^i)
 &= \psi (N, x^i) - \tilde{n} + \calO (\epsilon^2) \notag\\
 &= \psi (\tilde{N} + \tilde{n}, \tilde{x}^i) - \tilde{n}
 + \calO (\epsilon^2) \,.
 \end{align}
As is clear from Eq.~(\ref{GT-gamma}), $\gamma_{i j}$ is invariant at this order,
 \begin{align}
 \tilde{\gamma}_{i j} (\tilde{N}, \tilde{x}^i)
 &= \gamma_{i j} (N, x^i) + \calO (\epsilon^2) \notag\\
 &= \gamma_{i j} (\tilde{N} + \tilde{n}, \tilde{x}^i) + \calO (\epsilon^2) \,.
 \end{align}

Under the gauge transformation, the scalar field is essentially invariant 
 except for the change of the arguments,
\begin{gather}
 \tilde{\phi} (\tilde{N}, \tilde{x}^i) = \phi (N, x^i) \,.
\label{invscalar}
\end{gather}
Then, the leading order gauge transformation is
\begin{gather}
 \tilde{\phi} (\tilde{N}, \tilde{x}^i)
 = \phi (\tilde{N} + {}^{(0)}\tilde{n}, \tilde{x}^i) + \calO (\epsilon^2) \,.
\end{gather}

Here we note that what we call a gauge transformation should perhaps be
called a coordinate transformation as far as the change of time slicing is
concerned. For example, for a scalar quantity $Q$,
 the gauge transformation under the temporal shift ($t \to t + T$)
 is given by
 \begin{gather}
 \widetilde{\delta Q} (t, x^i) = \delta Q(t, x^i) -T\pa_t Q_0 \,,
\label{exam1}
 \end{gather}
 which is derived from the invariance of $Q$ as a scalar
 under the time coordinate transformation,
 \begin{gather}
 \tilde{Q}(\tilde{t}, x^i) = Q(t, x^i) \,.
\label{exam2}
 \end{gather}
In the context of gradient expansion, we define the nonlinear
gauge transformation of a scalar quantity by Eq.~(\ref{exam2})
not by Eq.~(\ref{exam1}),
since the gradient expansion is a completely nonlinear approach.
If we were to expand perturbatively the left-hand side of Eq.~(\ref{exam2}),
we would obtain an infinite number of terms, which is definitely something
to be avoided.
On the other hand, as we see shortly below,
to the accuracy of our interest, i.e. to $\calO(\epsilon^2)$,
the spatial coordinate transformation can be linearised. Hence we 
interpret it as the standard gauge transformation and compare the
quantities at the same coordinate values.

Once we have the gauge transformation rules for the metric components,
we can readily obtain those for the extrinsic curvature components,
 $K$ and $A_{i j}$. Since $A_{ij}$ vanishes at leading order,
we only have to consider the transformation of the expansion $K$.
From its definition, Eq.~(\ref{def-K}), and
using the transformation rules for the metric components derived above,
we find it is invariant at leading order, 
\begin{align}
 \tilde{K} (\tilde{N}, \tilde{x}^i) 
 &= K (\tilde{N} + {}^{(0)}\tilde{n}, x^i)
 + \calO (\epsilon^2) \,. 
\end{align}
Again, as discussed around Eq.~(\ref{exam2}), 
we note that $K$ is invariant only in the sense of
the nonlinear gauge transformation we have defined.

\subsection{Next-to-leading order gauge transformation}
Now we derive the gauge transformation at next-to-leading order.
To do so, we first have to fix the spatial coordinate transformation.
It is determined by Eq.~(\ref{GT-beta}), which results from the
requirement that the shift vector should vanish to 
$\calO(\epsilon^3)$. In terms of $\tilde{n}$, the gauge transformation
generator $\tilde{L}^i$ is given as
\begin{align} 
 \tilde{L}^i
 &= l^i (\tilde{x}^i)
 + \int_{\tilde{N}_0}^{\tilde{N}} d N'
 \frac{\alpha^2 \bigl( 1 + \pa_{N'} \tilde{n} \bigr)}{a^2 (N) e^{2 \psi}
 H^2 (N)} \gamma^{i j} \pa_{\tilde{j}} \tilde{n} + \calO (\epsilon^3) \notag\\
 &= l^i (\tilde{x}^i) + \int_{\tilde{N}_0}^{\tilde{N}} d N'
 \frac{\alpha^2 (N' + \tilde{n}, \tilde{x}^i)
 \bigl(1 + \pa_{N'} \tilde{n} \bigr)}{a^2 (N' + \tilde{n})
 e^{2 \psi (N' + \tilde{n}, \tilde{x}^i)} H^2 (N' + \tilde{n})}
 \gamma^{i j} (N' + \tilde{n}, \tilde{x}^i) \pa_{\tilde{j}} \tilde{n} 
 + \calO (\epsilon^3) \,,
\end{align}
 where $l^i$ is a time-independent spatial vector which represents
the remaining gauge degrees of freedom in the spatial coordinates
in the time-slice-orthogonal threading.

To consider the gauge transformation at next-to-leading order,
we expand the nonlinear gauge transformation generator $\tilde{n}$ in 
Eq.~(\ref{GT:n}) to the leading order term, the next-to-leading order 
term and so on,
 \begin{gather}
 \tilde{n} \equiv {}^{(0)}\tilde{n} + {}^{(2)}\tilde{n} + \calO (\epsilon^4) \,.
 \end{gather}
As for the spatial shift $\tilde{L}^i$, it is unnecessary to expand it,
since it is $\calO(\epsilon)$ already and the next-to-leading order term 
in $\tilde{L}^i$ is $\calO(\epsilon^3)$ which only affects the gauge 
transformation at $\calO (\epsilon^4)$.

First we consider the metric components.
From Eq.~(\ref{GT-alpha}), the next-to-leading order gauge
transformation for the lapse function is determined as
 \begin{align}   
 {}^{(2)}\tilde{\alpha}(\tilde{N}, \tilde{x}^i)
 &= \frac{\alpha H (\tilde{N})}{H ({}^{(0)}\tilde{N})} 
 \bigl(1 + \pa_{\tilde{N}}{}^{(0)}\tilde{n}\bigr)
 \left( \frac{{}^{(2)}\alpha}{\alpha}
 + {}^{(2)}\tilde{n} \frac{\pa_N \alpha}{\alpha}
 - {}^{(2)}\tilde{n} \frac{\pa_N H ({}^{(0)}\tilde{N})}{H ({}^{(0)}\tilde{N})}
 + \frac{\pa_{\tilde{N}} {}^{(2)}\tilde{n}}{1 + \pa_{\tilde{N}} {}^{(0)}\tilde{n}}
 \right) \notag\\
 & \qquad - \frac{1}{2} \frac{\alpha^3 H (\tilde{N})
 \bigl(1 + \pa_{\tilde{N}}{}^{(0)}\tilde{n}\bigr)}{a^2 ({}^{(0)}\tilde{N})
 e^{2 \psi} H^3 ({}^{(0)}\tilde{N})} \gamma^{i j} \pa_{\tilde{i}} {}^{(0)}\tilde{n}
 \pa_{{\tilde{j}}} {}^{(0)}\tilde{n}
 + \tilde{L}^i (\pa_{\tilde{i}} \alpha)
 \frac{H (\tilde{N})}{H({}^{(0)}\tilde{N})} 
 \bigl( 1 + \pa_{\tilde{N}}{}^{(0)}\tilde{n} \bigr) + \calO (\epsilon^4) \,,
 \end{align}
 where $\pa_{\tilde{i}} = \pa / \pa \tilde{x}^i$ and
 ${}^{(0)}\tilde{N} (N, x^i) \equiv N + {}^{(0)}\tilde{n} (N, x^i)$.
Note that all the quantities on the right-hand side are those
evaluated at $N = {}^{(0)}\tilde{N}$ and $x^i = \tilde{x}^i$.
Taking the determinant of Eq.~(\ref{GT-gamma}), 
\begin{gather}
 a^6 (\tilde{N}) e^{6 \tilde{\psi}} 
 = a^6 (N) e^{6 \psi} \bigl( 1 + 2 \pa_{\tilde{i}} \tilde{L}^i \bigr)
 - a^4 (N) e^{4 \psi} \frac{\alpha^2}{H^2 (N)}
 \gamma^{i j} \pa_{\tilde{i}} \tilde{n} \pa_{\tilde{j}} \tilde{n}
 + \calO (\epsilon^4) \,,
\end{gather}
 we obtain the gauge transformation for ${}^{(2)}\psi$,
\begin{align}   
 {}^{(2)}\tilde{\psi} (\tilde{N}, \tilde{x}^i) 
 &= {}^{(2)}\psi - {}^{(2)}\tilde{n} + {}^{(2)}\tilde{n} \pa_N \psi 
 + \tilde{L}^i \pa_{\tilde{i}} \psi + \frac{1}{3} \pa_{\tilde{i}} \tilde{L}^i 
 - \frac{1}{6} \frac{\alpha^2}{a^2 ({}^{(0)}\tilde{N}) e^{2 \psi}
 H^2 ({}^{(0)}\tilde{N})} \gamma^{i j} \pa_{\tilde{i}} {}^{(0)}\tilde{n}
 \pa_{{\tilde{j}}}{}^{(0)}\tilde{n} + \calO (\epsilon^4) \,.
\end{align}
Then using Eq.~(\ref{GT-gamma}) again, 
the gauge transformation for $\gamma_{i j}$ is obtained as
\begin{align}  
{}^{(2)}\tilde{\gamma}_{i j}(\tilde{N}, \tilde{x}^i)
 &= {}^{(2)}\gamma_{i j}+ {}^{(2)}\tilde{n} \pa_{\tilde{N}}\gamma_{i j}
 + \tilde{L}^k \pa_{\tilde{k}} \gamma_{i j}
 + \gamma_{j k} \pa_{\tilde{i}} \tilde{L}^k
 + \gamma_{i k} \pa_{\tilde{j}} \tilde{L}^k
 - \frac{2}{3} \pa_{\tilde{i}} \tilde{L}^i \notag\\
 & \qquad
 - \frac{\alpha^2}{a^2 ({}^{(0)}\tilde{N}) e^{2 \psi} H^2 ({}^{(0)}\tilde{N})}
 \left( \pa_{\tilde{i}} {}^{(0)}\tilde{n} \pa_{\tilde{j}} {}^{(0)}\tilde{n}
 - \frac{1}{3} \gamma^{k l} \pa_{\tilde{k}}{}^{(0)}\tilde{n}
 \pa_{\tilde{l}}{}^{(0)}\tilde{n} \gamma_{i j} \right) + \calO (\epsilon^4) \,.
\end{align}

Next we consider the extrinsic curvature.
At next-to-leading order, Eq.~(\ref{def-K}) gives
\begin{align}
 {}^{(2)}\tilde{K} (\tilde{N}, \tilde{x}^i) 
 &= {}^{(2)}K + {}^{(2)}\tilde{n} \pa_N {}^{(0)}K
 + \tilde{L}^i \pa_{\tilde{i}}{}^{(0)}K
 + \frac{3 H ({}^{(0)}\tilde{N})}{\bigl(1 + \pa_{\tilde{N}}{}^{(0)}n\bigr) \alpha}
 \left( \pa_{\tilde{N}}{}^{(0)}\tilde{n} \frac{{}^{(2)}\alpha}{\alpha}
 - \frac{\pa_{\tilde{N}} {}^{(2)}\tilde{n}}{1 + \pa_{\tilde{N}}{}^{(0)}\tilde{n}}
 \right) \notag\\
 & \qquad 
 + \frac{3 \alpha \gamma^{i j} \pa_{\tilde{i}} {}^{(0)}\tilde{n}
 \pa_{{\tilde{j}}} {}^{(0)}\tilde{n}}{2 \bigl(1 + \pa_{\tilde{N}} {}^{(0)}n\bigr)
 a^2 ({}^{(0)}\tilde{N}) e^{2 \psi} H ({}^{(0)}\tilde{N})}
 + \frac{\pa_{\tilde{i}}{}^{(0)}\tilde{n} \pa_{{\tilde{j}}}{}^{(0)}\tilde{n}}
 {2 \alpha a^2 ({}^{(0)}\tilde{N}) H ({}^{(0)}\tilde{N})} 
 \pa_N \left( \frac{\alpha^2}{e^{2 \psi}} \gamma^{i j} \right) \notag\\
 & \qquad 
 + \frac{3 H ({}^{(0)}\tilde{N})}{\bigl(1 + \pa_{\tilde{N}} {}^{(0)}n\bigr) \alpha}
 \left[ \pa_{\tilde{N}} {}^{(2)}\tilde{n} \bigl(1 - \pa_N \psi\bigr) 
 - \bigl(\pa_{\tilde{N}} \tilde{L}^i\bigr) \pa_{\tilde{i}} \psi
 - \frac{1}{3} \pa_{\tilde{i}} \pa_{\tilde{N}} \tilde{L}^i \right] \notag\\
 & \qquad 
 + \frac{H ({}^{(0)}\tilde{N}) \alpha \gamma^{i j}}{2
 \bigl(1 + \pa_{\tilde{N}}{}^{(0)}n\bigr) e^{2 \psi}}
 \pa_{\tilde{N}} \left( \frac{\pa_{\tilde{i}} {}^{(0)}\tilde{n}
 \pa_{{\tilde{j}}} {}^{(0)}\tilde{n}}{a^2({}^{(0)}\tilde{N}) H^2({}^{(0)}\tilde{N})}
 \right) + \calO (\epsilon^4) \,.
\end{align}
And also Eq.~(\ref{eq-gamma}) determines the gauge transformation
 for $A_{i j}$ as
\begin{align}
 \tilde{A}_{i j} (\tilde{N}, \tilde{x}^i)
 &= A_{i j} - \frac{H ({}^{(0)}\tilde{N})}{2
 \bigl(1 + \pa_{\tilde{N}} {}^{(0)}\tilde{n}\bigr) \alpha}
 \Bigl( \pa_{\tilde{N}} \tilde{L}^k \pa_{\tilde{k}} \gamma_{i j}
 + \gamma_{j k} \pa_{\tilde{i}} \pa_{\tilde{N}} \tilde{L}^k
 + \gamma_{i k} \pa_{\tilde{j}} \pa_{\tilde{N}} \tilde{L}^k \Bigr)^{T F} \notag\\
 & \qquad
 - \frac{H ({}^{(0)}\tilde{N})}{2 \bigl(1 + \pa_{\tilde{N}} {}^{(0)}\tilde{n}\bigr)
 \alpha} \pa_{\tilde{N}} \left[ \frac{\alpha^2}{a^2 ({}^{(0)}\tilde{N}) e^{2 \psi}
 H^2 ({}^{(0)}\tilde{N})} \Bigl( \pa_{\tilde{i}}{}^{(0)}\tilde{n}
 \pa_{\tilde{j}} {}^{(0)}\tilde{n} \Bigr)^{T F} \right] 
 + \calO (\epsilon^4) \,.
\end{align}

Finally, the next-to-leading order gauge transformation for the
scalar field is given from Eq.~(\ref{invscalar}) as
\begin{gather}
 {}^{(2)}\tilde{\phi} (\tilde{N}, \tilde{x}^i)
 = {}^{(2)}\phi (\tilde{N} +{}^{(0)}\tilde{n}, \tilde{x}^i)
 + {}^{(2)}\tilde{n} \pa_N{}^{(0)}\phi(\tilde{N} +{}^{(0)}\tilde{n}, \tilde{x}^i)
 + \tilde{L}^i \pa_{\tilde{i}} {}^{(0)}\phi(\tilde{N} +{}^{(0)}\tilde{n},
 \tilde{x}^i) + \calO (\epsilon^2) \,.
\end{gather}

To conclude this Appendix, let us
summarise the nonlinear gauge transformation derived above.
The leading order transformation rules are
 \begin{align}
 \tilde{\alpha} (\tilde{N}, \tilde{x}^i)
 &= \frac{H (\tilde{N})}{H ({}^{(0)}\tilde{N})}
 \bigl(1 + \pa_{\tilde{N}}{}^{(0)}\tilde{n}\bigr)
 ~ \alpha + \calO (\epsilon^2) \,, \\
 \tilde{\psi} (\tilde{N}, \tilde{x}^i)
 &= \psi - {}^{(0)}\tilde{n} + \calO (\epsilon^2) \,, 
\label{GT-psi-0}\\
 \tilde{\gamma}_{i j} (\tilde{N}, \tilde{x}^i)
 &= \gamma_{i j} + \calO (\epsilon^2) \,, 
\label{GT-gamma-0}\\
 \tilde{K} (\tilde{N}, \tilde{x}^i)
 &= K + \calO (\epsilon^2) \,, \label{gauT-K}
\\
 \tilde{\phi} (\tilde{N}, \tilde{x}^i)
 &= \phi + \calO (\epsilon^2) \,.
 \end{align}
The next-to-leading order transformation rules are
 \begin{align}   
{}^{(2)} \tilde{\alpha}(\tilde{N}, \tilde{x}^i)
 &= \frac{\alpha H (\tilde{N})}{H ({}^{(0)}\tilde{N})} 
 \bigl(1 + \pa_{\tilde{N}}{}^{(0)}\tilde{n}\bigr)
 \left( \frac{{}^{(2)}\alpha}{\alpha}
 + {}^{(2)}\tilde{n} \frac{\pa_N \alpha}{\alpha}
 - {}^{(2)}\tilde{n} \frac{\pa_N H ({}^{(0)}\tilde{N})}{H ({}^{(0)}\tilde{N})}
 + \frac{\pa_{\tilde{N}} {}^{(2)}\tilde{n}}{1 + \pa_{\tilde{N}} {}^{(0)}\tilde{n}}
 \right) \notag\\
 & \qquad - \frac{1}{2} \frac{\alpha^3 H (\tilde{N})
 \bigl(1 + \pa_{\tilde{N}} {}^{(0)}\tilde{n}\bigr)}{a^2 ({}^{(0)}\tilde{N})
 e^{2 \psi} H^3 ({}^{(0)}\tilde{N})} \gamma^{i j} \pa_{\tilde{i}}{}^{(0)}\tilde{n}
 \pa_{{\tilde{j}}} {}^{(0)}\tilde{n}
 + \tilde{L}^i \bigl(\pa_{\tilde{i}} \alpha\bigr)
 \frac{H (\tilde{N})}{H ({}^{(0)}\tilde{N})} 
 \bigl(1 + \pa_{\tilde{N}}{}^{(0)}\tilde{n}\bigr) + \calO (\epsilon^4) \,, \\
 {}^{(2)}\tilde{\psi} (\tilde{N}, \tilde{x}^i) 
 &= {}^{(2)}\psi - {}^{(2)}\tilde{n} + {}^{(2)}\tilde{n} \pa_N \psi 
 + \tilde{L}^i \pa_{\tilde{i}} \psi + \frac{1}{3} \pa_{\tilde{i}} \tilde{L}^i 
 - \frac{1}{6} \frac{\alpha^2 \gamma^{i j} \pa_{\tilde{i}} {}^{(0)}\tilde{n}
 \pa_{{\tilde{j}}}{}^{(0)}\tilde{n}}{a^2 ({}^{(0)}\tilde{N}) e^{2 \psi}
 H^2 ({}^{(0)}\tilde{N})} + \calO (\epsilon^4) \,,
\label{GT-psi-2} \\
 {}^{(2)}\tilde{\gamma}_{i j}(\tilde{N}, \tilde{x}^i) 
 &= {}^{(2)}\gamma_{i j} + {}^{(2)}\tilde{n} \pa_{\tilde{N}}\gamma_{i j}
 + \tilde{L}^k \pa_{\tilde{k}} \gamma_{i j}
 + \gamma_{j k} \pa_{\tilde{i}} \tilde{L}^k
 + \gamma_{i k} \pa_{\tilde{j}} \tilde{L}^k
 - \frac{2}{3} \pa_{\tilde{k}} \tilde{L}^k \gamma_{i j} \notag\\
 & \qquad
 - \frac{\alpha^2}{a^2 ({}^{(0)}\tilde{N}) e^{2 \psi} H^2 ({}^{(0)}\tilde{N})}
 \left( \pa_{\tilde{i}} {}^{(0)}\tilde{n} \pa_{\tilde{j}} {}^{(0)}\tilde{n}
 - \frac{1}{3} \gamma^{k l} \pa_{\tilde{k}}{}^{(0)}\tilde{n}
 \pa_{\tilde{l}}{}^{(0)}\tilde{n} \gamma_{i j} \right) + \calO (\epsilon^4) \,,
\label{GT-gamma-2}
\end{align}
 and
\begin{align}
 {}^{(2)}\tilde{K} (\tilde{N}, \tilde{x}^i) 
 &= {}^{(2)}K + {}^{(2)}\tilde{n} \pa_N {}^{(0)}K + \tilde{L}^i \pa_{\tilde{i}}{}^{(0)}K
 + \frac{3 H ({}^{(0)}\tilde{N})}{\bigl(1 + \pa_{\tilde{N}}{}^{(0)}n\bigr) \alpha}
 \left( \pa_{\tilde{N}}{}^{(0)}\tilde{n} \frac{{}^{(2)}\alpha}{\alpha}
 - \frac{\pa_{\tilde{N}} {}^{(2)}\tilde{n}}{1 + \pa_{\tilde{N}} {}^{(0)}\tilde{n}}
 \right) \notag\\
 & \qquad 
 + \frac{3 \alpha \gamma^{i j} \pa_{\tilde{i}}{}^{(0)}\tilde{n}
 \pa_{{\tilde{j}}}{}^{(0)}\tilde{n}}{2 \bigl(1 + \pa_{\tilde{N}} {}^{(0)}n\bigr)
 a^2 ({}^{(0)}\tilde{N}) e^{2 \psi} H ({}^{(0)}\tilde{N})}
 + \frac{\pa_{\tilde{i}} {}^{(0)}\tilde{n} \pa_{{\tilde{j}}} {}^{(0)}\tilde{n}}{2
 \alpha a^2 ({}^{(0)}\tilde{N}) H ({}^{(0)}\tilde{N})} 
 \pa_N \left( \frac{\alpha^2}{e^{2 \psi}} \gamma^{i j} \right) \notag\\
 & \qquad 
 + \frac{3 H ({}^{(0)}\tilde{N})}{\bigl(1 + \pa_{\tilde{N}}{}^{(0)}n\bigr) \alpha}
 \left[ \pa_{\tilde{N}} {}^{(2)}\tilde{n} \bigl(1 - \pa_N \psi\bigr) 
 - \bigl(\pa_{\tilde{N}} \tilde{L}^i\bigr) \pa_{\tilde{i}} \psi
 - \frac{1}{3} \pa_{\tilde{i}} \pa_{\tilde{N}} \tilde{L}^i \right] \notag\\
 & \qquad 
 + \frac{H ({}^{(0)}\tilde{N}) \alpha \gamma^{i j}}{2
 \bigl(1 + \pa_{\tilde{N}}{}^{(0)}n\bigr) e^{2 \psi}}
 \pa_{\tilde{N}} \left( \frac{\pa_{\tilde{i}}{}^{(0)}\tilde{n}
 \pa_{{\tilde{j}}} {}^{(0)}\tilde{n}}{a^2({}^{(0)}\tilde{N}) H^2({}^{(0)}\tilde{N})}
 \right) + \calO (\epsilon^4) \,,
 \label{GT-K-2} \\
 \tilde{A}_{i j} (\tilde{N}, \tilde{x}^i)
 &= A_{i j} - \frac{H ({}^{(0)}\tilde{N})}{2
 \bigl(1 + \pa_{\tilde{N}}{}^{(0)}\tilde{n}\bigr) \alpha}
 \Bigl( \pa_{\tilde{N}} \tilde{L}^k \pa_{\tilde{k}} \gamma_{i j}
 + \gamma_{j k} \pa_{\tilde{i}} \pa_{\tilde{N}} \tilde{L}^k
 + \gamma_{i k} \pa_{\tilde{j}} \pa_{\tilde{N}} \tilde{L}^k \Bigr)^{T F} \notag\\
 & \qquad
 - \frac{H ({}^{(0)}\tilde{N})}{2 \bigl(1 + \pa_{\tilde{N}} {}^{(0)}\tilde{n}\bigr)
 \alpha} \pa_{\tilde{N}} \left[ \frac{\alpha^2}{a^2 ({}^{(0)}\tilde{N}) e^{2 \psi}
 H^2 ({}^{(0)}\tilde{N})} \Bigl( \pa_{\tilde{i}} {}^{(0)}\tilde{n}
 \pa_{\tilde{j}} {}^{(0)}\tilde{n} \Bigr)^{T F} \right] 
 + \calO (\epsilon^4) \,, \\
 {}^{(2)}\tilde{\phi} (\tilde{N}, \tilde{x}^i)
 &= {}^{(2)}\phi + {}^{(2)}\tilde{n} \pa_N{}^{(0)}\phi
 + \tilde{L}^i \pa_{\tilde{i}} {}^{(0)}\phi + \calO (\epsilon^2) \,,
\label{GT-phi-2}
\end{align}
 where $\tilde{L}^i$ is given by
\begin{align} 
 \tilde{L}^i
 &= l^i (\tilde{x}^i) + \int_{\tilde{N}_0}^{\tilde{N}} d N'
 \frac{\alpha^2 \bigl(1 + \pa_{N'} \tilde{n}\bigr)}{ a^2({}^{(0)}\tilde{N})
 e^{2 \psi} H^2 ({}^{(0)}\tilde{N})}\gamma^{i j}
 \pa_{\tilde{j}} \tilde{n} + \calO (\epsilon^3) \,.
\label{tildeL}
\end{align}

%%%%%%%%%%%%%%%%%%%%%%%%%%%%%%%%%%%%%%%%%%%%%%%%%%%%%%%%%%%%%%%%%%%%%%%%%%
\section{Consistency checking}
\label{app:verify}

In this Appendix, we apply our formalism to the case of a single scalar
field and verify the consistency.
One way to do this is just to compare the obtained solutions
 in this formalism with the results obtained previously,
for example, in~\cite{Takamizu:2010xy}.
Here, however, we take another, direct way. Namely, we consider the solution
in the comoving gauge and transform it to the one in the $\calN$ gauge.
Then we insert thus derived solutions for the metric and the scalar field
into the basic equations in the $\calN$ gauge and show that they
are satisfied.

%%%%%%%%%%%%%%%%%%%%%%%%%%%%%%%%%%%%%%%%%%%%%%%%%%%%%%%%%%%%%%%%%%%%%%%%%%
\subsection{Solution in the uniform $\calN$ gauge}

First we construct the solution in the $\calN$ gauge 
by transforming it from the one in the comoving gauge.
As we have seen in Appendix~\ref{app:coincide}, 
 the leading order solution in the comoving gauge is
\begin{gather}
 {}^{(0)}\alpha_c = f_\alpha (N) \,, \qquad {}^{(0)}\psi_c = \calC (x^i) \,, \qquad
{}^{(0)}\gamma_{i j c} = \calC_{i j} (x^i) \,, \qquad K_c = f_K (N) \,, \qquad
 \phi_c = f_\phi (N) \,.
\label{comsol}
\end{gather}
Here $f_\alpha$, $f_K$ and $f_\phi$ are functions of only time and
 have no spatial dependence.
On the other hand, $\calC$ and $\calC_{i j}$ are functions of only spatial 
 coordinates and have no time dependence.
Because of the definition of $K$, Eq.~(\ref{def-K}), the functions
$f_K$ are $f_\alpha$ are related as
 \begin{align}
 f_K = \frac{3 H}{f_\alpha} \,.
 \end{align}
At next-to-leading order, however, all the quantities listed in Eq.~(\ref{comsol})
become space-time dependent except for the scalar field ${}^{(2)}\phi$, 
which is by definition a function of only time. Hence without loss of
generality, we may absorb it in the leading order scalar field ${}^{(0)}\phi$
and set ${}^{(2)}\phi=0$.

To perform the transformation from the comoving slicing to the 
uniform $\calN$ slicing, $N=\tilde{N}+\tilde{n} (\tilde{N}, \tilde{x}^i)$,
one needs to determine $\tilde{n}$. In the uniform $\calN$ slicing, 
$\psi$ is time-independent by definition, Eq.~(\ref{def-calN}).
Considering the transformation of $\psi$, this gives at leading order,
 \begin{align}
 \psi_\calN (\tilde{x}^i)
 &= \calC(x^i) -{}^{(0)}\tilde{n} + \calO (\epsilon^2) \quad \to \quad
 {}^{(0)}\tilde{n} = \calC(\tilde{x}^i) - \psi_\calN (\tilde{x}^i)
 + \calO (\epsilon^2) \,.
\label{appV:n-0}
 \end{align}
At the next-leading-order Eq.~(\ref{GT-psi-2}) gives
 \begin{align}
 0
 &= {}^{(2)}\psi_c - {}^{(2)}\tilde{n} + \tilde{L}^i \pa_{\tilde{i}} \calC
 + \frac{1}{3} \pa_{\tilde{i}} \tilde{L}^i 
 - \frac{1}{6} \frac{f_\alpha^2 \calC^{i j} \pa_{\tilde{i}} {}^{(0)}\tilde{n}
 \pa_{\tilde{j}}{}^{(0)}\tilde{n}}{a^2 ({}^{(0)}\tilde{N}) e^{2 \calC}
 H^2 ({}^{(0)}\tilde{N})} + \calO (\epsilon^4) \notag\\
 &\to \quad
 {}^{(2)}\tilde{n}
 = {}^{(2)}\psi_c + \tilde{L}^i \pa_{\tilde{i}} \calC
 + \frac{1}{3} \pa_{\tilde{i}} \tilde{L}^i 
 - \frac{1}{6} \frac{f_\alpha^2 \calC^{i j} \pa_{\tilde{i}} {}^{(0)}\tilde{n}
 \pa_{\tilde{j}}{}^{(0)}\tilde{n}}{a^2 ({}^{(0)}\tilde{N}) e^{2 \calC}
 H^2 ({}^{(0)}\tilde{N})} + \calO (\epsilon^4) \,, 
\label{appV:n-2}
 \end{align}
where $\tilde{L}^i$ is given by Eq.~(\ref{tildeL}), which is derived
from the time-slice-orthogonal threading condition,
and the fact that ${}^{(2)}\psi_\calN=0$ is used since it may be 
absorbed into ${}^{(0)}\psi_\calN$.
Note that from Eq.~(\ref{appV:n-0}), ${}^{(0)} \tilde{n}$ is independent
of time which significantly simplifies the analysis below.

Performing the gauge transformation, we we obtain at leading order, 
 \begin{align}
 \alpha_\calN (\tilde{N}, \tilde{x}^i)
 &= \frac{H (\tilde{N})}{H ({}^{(0)}\tilde{N})} f_\alpha ({}^{(0)}\tilde{N})
 + \calO (\epsilon^2) \,, \\
 \gamma_{i j \calN} (\tilde{N}, \tilde{x}^i)
 &= \calC_{i j} + \calO (\epsilon^2) \,, \\
 K_\calN (\tilde{N}, \tilde{x}^i)
 &= f_K ({}^{(0)}\tilde{N}) + \calO (\epsilon^2) \,, \\
 \phi_\calN (\tilde{N}, \tilde{x}^i)
 &= f_\phi ({}^{(0)}\tilde{N}) + \calO (\epsilon^2) \,,
 \end{align}
and at the next-to-leading order,
 \begin{align}   
 {}^{(2)} \alpha_\calN(\tilde{N}, \tilde{x}^i)
 &= \frac{f_\alpha H (\tilde{N})}{H ({}^{(0)}\tilde{N})} 
 \left( \frac{{}^{(2)}\alpha_c}{f_\alpha}
 + {}^{(2)}\tilde{n} \frac{\pa_N f_\alpha}{f_\alpha}
 - {}^{(2)}\tilde{n} \frac{\pa_N H ({}^{(0)}\tilde{N})}{H ({}^{(0)}\tilde{N})}
 + \pa_{\tilde{N}} {}^{(2)}\tilde{n} \right) 
 - \frac{1}{2} \frac{f_\alpha^3 \calC^{i j} \pa_{\tilde{i}}{}^{(0)}\tilde{n}
 \pa_{\tilde{j}}{}^{(0)}\tilde{n} H (\tilde{N})}{a^2 ({}^{(0)}\tilde{N})
 e^{2 \calC} H^3 ({}^{(0)}\tilde{N})} + \calO (\epsilon^4) \,,
\label{appV:GT-alpha} \\
 {}^{(2)}\gamma_{i j \calN}(\tilde{N}, \tilde{x}^i) 
 &= {}^{(2)}\gamma_{i j c} + \tilde{L}^k \pa_{\tilde{k}} \calC_{i j}
 + \calC_{j k} \pa_{\tilde{i}} \tilde{L}^k + \calC_{i k} \pa_{\tilde{j}} \tilde{L}^k
 - \frac{2}{3} \pa_{\tilde{k}} \tilde{L}^k \calC_{i j} \notag\\
 & \qquad
 - \frac{f_\alpha^2}{a^2 ({}^{(0)}\tilde{N}) e^{2 \calC} H^2 ({}^{(0)}\tilde{N})}
 \left( \pa_{\tilde{i}} {}^{(0)}\tilde{n} \pa_{\tilde{j}}{}^{(0)}\tilde{n}
 - \frac{1}{3} \calC^{k l} \pa_{\tilde{k}} {}^{(0)}\tilde{n}
 \pa_{\tilde{l}} {}^{(0)}\tilde{n} \calC_{i j} \right) + \calO (\epsilon^4) \,,
\end{align}
and
\begin{align}
{}^{(2)}K_\calN(\tilde{N}, \tilde{x}^i) 
 &= {}^{(2)}K_c + {}^{(2)}\tilde{n} \pa_N f_K + \left( \frac{3}{2} f_\alpha
 + \pa_N f_\alpha \right) \frac{\calC^{i j} \pa_{\tilde{i}} {}^{(0)}\tilde{n}
 \pa_{\tilde{j}} {}^{(0)}\tilde{n}}{a^2 ({}^{(0)}\tilde{N}) e^{2 \calC}
 H ({}^{(0)}\tilde{N})} 
 - \frac{3 H ({}^{(0)}\tilde{N})}{f_\alpha}
 \left[ \bigl(\pa_{\tilde{N}} \tilde{L}^i\bigr) \pa_{\tilde{i}} \calC
 + \frac{1}{3} \pa_{\tilde{i}} \pa_{\tilde{N}} \tilde{L}^i \right] \notag\\
 & \qquad 
 + \frac{H ({}^{(0)}\tilde{N}) f_\alpha \calC^{i j} \pa_{\tilde{i}}
 {}^{(0)}\tilde{n}
 \pa_{\tilde{j}}{}^{(0)}\tilde{n}}{2 e^{2 \calC}}
 \pa_{\tilde{N}} \left( \frac{1}{a^2
 ({}^{(0)}\tilde{N}) H^2 ({}^{(0)}\tilde{N})} \right) + \calO (\epsilon^4) \,,
\label{appV:GT-K} \\
 A_{i j \calN} (\tilde{N}, \tilde{x}^i)
 &= A_{i j c} - \frac{H ({}^{(0)}\tilde{N})}{2 f_\alpha}
 \Bigl( \pa_{\tilde{N}} \tilde{L}^k \pa_{\tilde{k}} \calC_{i j}
 + \calC_{j k} \pa_{\tilde{i}} \pa_{\tilde{N}} \tilde{L}^k
 + \calC_{i k} \pa_{\tilde{j}} \pa_{\tilde{N}} \tilde{L}^k \Bigr)^{T F} \notag\\
 & \qquad
 + \frac{H ({}^{(0)}\tilde{N})}{2 f_\alpha e^{2 \calC}} \Bigl(
 \pa_{\tilde{i}} {}^{(0)}\tilde{n} \pa_{\tilde{j}}{}^{(0)}\tilde{n}\Bigr)^{T F}
 \pa_{\tilde{N}} \left( \frac{f_\alpha^2}{a^2({}^{(0)}\tilde{N})
 H^2({}^{(0)}\tilde{N})} \right) + \calO (\epsilon^4) \,,
\label{appV:GT-Aij} \\
 {}^{(2)}\phi_\calN(\tilde{N}, \tilde{x}^i)
 &= {}^{(2)}\tilde{n} \pa_N f_\phi + \calO (\epsilon^4) \,,
\end{align}
 where $\pa_{\tilde{i}} = \pa / \pa \tilde{x}^i$ and $\tilde{L}^i$ is
given explicitly by
\begin{align} 
 \tilde{L}^i
 &= l^i (\tilde{x}^i) +
 \frac{\calC^{i j} \pa_{\tilde{j}}{}^{(0)}\tilde{n}}{e^{2 \calC}}
 \int_{\tilde{N}_0}^{\tilde{N}} d N' \frac{f_\alpha^2}{a^2({}^{(0)}\tilde{N})
 H^2({}^{(0)}\tilde{N})} + \calO (\epsilon^3) \,.
\end{align}

After some manipulations, Eqs.~(\ref{appV:GT-alpha}),
 (\ref{appV:GT-K}) and (\ref{appV:GT-Aij}) may be re-expressed as
\begin{align}  
 \frac{\alpha_\calN}{\alpha_\calN^{(0)}} 
 - \frac{\alpha_c}{f_\alpha}
 &= \pa_{\tilde{N}} {}^{(2)}\tilde{n} 
 - {}^{(2)}\tilde{n} \pa_N \log f_K ({}^{(0)}\tilde{N})
 - \frac{9}{2} \frac{D^{\tilde{i}} {}^{(0)}\tilde{n} D_{\tilde{i}}
 {}^{(0)}\tilde{n}}{a^2 ({}^{(0)}\tilde{N}) e^{2 \calC} f_K^2 ({}^{(0)}\tilde{N})} \,,
\label{appV:N:sol-alpha} \\
 K_\calN - K_c
 &= {}^{(2)}\tilde{n} \pa_N f_K ({}^{(0)}\tilde{N}) 
 + 3 \left( \frac{1}{2} + \pa_N \log f_K ({}^{(0)}\tilde{N}) \right)
 \frac{D^{\tilde{i}} {}^{(0)}\tilde{n} \pa_{\tilde{j}} {}^{(0)}\tilde{n}}{a^2
 ({}^{(0)}\tilde{N}) e^{2 \calC} f_K ({}^{(0)}\tilde{N})}
 - 3 \frac{D^2\,{}^{(0)}\tilde{n} + D^{\tilde{i}} {}^{(0)}\tilde{n}
 D_{\tilde{i}} \calC}{a^2 ({}^{(0)}\tilde{N}) e^{2 \calC} f_K ({}^{(0)}\tilde{N})} \,,
\label{appV:N:sol-K} \\
 A_{i j \calN} - A_{i j c}
 &= - \frac{3}{a^2 ({}^{(0)}\tilde{N}) e^{2 \calC} f_K ({}^{(0)}\tilde{N})}
 \Bigl[ D_{\tilde{i}} D_{\tilde{j}} {}^{(0)}\tilde{n}
 - \Bigl( D_{\tilde{i}} {}^{(0)}\tilde{n} D_{\tilde{j}} \calC
 + D_{\tilde{j}} {}^{(0)}\tilde{n} D_{\tilde{i}} \calC \Bigr) \Bigr]^{T F} \notag\\
 & \qquad
  - \frac{3}{a^2 ({}^{(0)}\tilde{N}) e^{2 \calC} f_K ({}^{(0)}\tilde{N})}
 \Bigl( 1 - \pa_N \log f_K ({}^{(0)}\tilde{N}) \Bigr)
 \Bigl( D_{\tilde{i}} {}^{(0)}\tilde{n} D_{\tilde{j}} {}^{(0)}\tilde{n} \Bigr)^{T F} \,,
\label{appV:N:sol-Aij}
\end{align}
 where $D_{\tilde{i}}$ is a covariant derivative with respect
 to ${}^{(0)}\gamma_{i j \calN}$ (or equivalently to $\calC_{i j}$).
These expressions are convenient for later use.

%%%%%%%%%%%%%%%%%%%%%%%%%%%%%%%%%%%%%%%%%%%%%%%%%%%%%%%%%%%%%%%%%%%%%%%%%%
\subsection{Consistency check for $A_{i j}$}

Here we explicitly show that the solution for $A_{i j}$
obtained in the previous subsection satisfies the evolution equation
 in the $\calN$ gauge. This verifies the consistency of the derived 
gauge transformation.

In the $\calN$ gauge, the evolution equations for $A_{i j}$ is
\begin{align}
 \pa_{\tilde{N}} A_{i j \calN} 
 &= 3 A_{i j \calN}
 + \frac{3}{a^2 (\tilde{N}) e^{2\psi_\calN} K_\calN}
 \Bigl( R_{i j} + D_{\tilde{i}} \psi_\calN D_{\tilde{j}} \psi_\calN
 - D_{\tilde{i}} D_{\tilde{j}} \psi_\calN \Bigr)^{T F} \notag\\
 & \qquad
 - \frac{3}{a^2 (\tilde{N}) e^{2\psi_\calN} K_\calN}
 \left[ K_\calN D_{\tilde{i}} D_{\tilde{j}} \left( \frac{1}{K_\calN} \right)
 + D_{\tilde{i}} \log K_\calN D_{\tilde{j}} \psi_\calN
 + D_{\tilde{j}} \log K_\calN D_{\tilde{i}} \psi_\calN
 + D_{\tilde{i}} \phi_\calN D_{\tilde{j}} \phi_\calN \right]^{TF} \,,
\label{appV:N:eq-Aij}
\end{align}
while in the comoving gauge it is
\begin{align}
 \pa_N A_{i j c}
 &= 3 A_{i j c} + \frac{3}{a^2 (N) e^{2 \calC} f_K (N)}
 \Bigl( R_{i j} + D_i \calC D_j \calC - D_i D_j \calC \Bigr)^{TF} \,.
\label{appV:c:eq-Aij}
\end{align}

We rewrite the second line inside the brackets of Eq.~ (\ref{appV:N:eq-Aij}) 
as
 \begin{align}  
 (\ma{2nd ~ line}) 
 &= \bigl(\pa_{\tilde{N}} \log f_K\bigr) ({}^{(0)}\tilde{N})
 \Bigl[ D_{\tilde{i}} D_{\tilde{j}} {}^{(0)}\tilde{n}
 - \Bigl( D_{\tilde{i}} {}^{(0)}\tilde{n} D_{\tilde{j}} \psi_\calN
 + D_{\tilde{j}} {}^{(0)}\tilde{n} D_{\tilde{i}} \psi_\calN \Bigr) \Bigr]
 \notag\\
 & \qquad
 + \Bigl[ \pa_{\tilde{N}}^2 \log f_K - \bigl(\pa_{\tilde{N}} \log f_K\bigr)^2
 - 2 \pa_{\tilde{N}} \log f_K \Bigr] ({}^{(0)}\tilde{N})
 D_{\tilde{i}} {}^{(0)}\tilde{n} D_{\tilde{j}} {}^{(0)}\tilde{n} \,,
\label{appV:2nd-Aij}
 \end{align}
 where we have used the leading order Einstein equation,
\begin{gather}
 2 \pa_{\tilde{N}} \log K_\calN = \bigl(\pa_{\tilde{N}} \phi_\calN\bigr)^2 \,.
\label{appV:BG:eq}
\end{gather}
By subtracting Eq.~(\ref{appV:c:eq-Aij}) from Eq.~(\ref{appV:N:eq-Aij}),
 we find with the aide of Eq.~(\ref{appV:2nd-Aij}),
\begin{align}
 \pa_{\tilde{N}} A_{i j} \Bigr|^\calN_c
 &= 3 A_{i j} \Bigr|^\calN_c
 + 3 \frac{1 + \pa_{\tilde{N}}\log f_K ({}^{(0)}\tilde{N})}
{a^2 ({}^{(0)}\tilde{N}) e^{2 \calC} f_K ({}^{(0)}\tilde{N})}
 \Bigl[ D_{\tilde{i}} D_{\tilde{j}} {}^{(0)}\tilde{n}
 - \Bigl( D_{\tilde{i}} {}^{(0)}\tilde{n} D_{\tilde{j}}\calC
 + D_{\tilde{j}} {}^{(0)}\tilde{n} D_{\tilde{i}} \calC \Bigr) \Bigr]^{TF} 
\notag\\
 & \qquad
 + \frac{3}{a^2 ({}^{(0)}\tilde{N}) e^{2 \calC} f_K ({}^{(0)}\tilde{N})} 
 \Bigl[ 1 + \pa_N^2 \log f_K - \bigl(\pa_N \log f_K\bigr)^2 \Bigr]
 ({}^{(0)}\tilde{N}) \Bigl( D_{\tilde{i}} {}^{(0)}\tilde{n} 
D_{\tilde{j}} {}^{(0)}\tilde{n} \Bigr)^{TF} \,, 
\label{appV:consis-Aij}
\end{align}
 where $Q \bigr|^\calN_c = Q_\calN - Q_c$.

On the other hand, taking the time derivative of Eq.~(\ref{appV:N:sol-Aij}),
which we have obtained by the gauge transformation,
we obtain
\begin{align}
 \pa_{\tilde{N}} A_{i j} \Bigr|^\calN_c
 &= 3 A_{i j} \Bigr|^\calN_c
 + 3 \frac{1 + \pa_{\tilde{N}} \log f_K ({}^{(0)}\tilde{N})}
{a^2 ({}^{(0)}\tilde{N}) e^{2 \calC} f_K ({}^{(0)}\tilde{N})}
 \Bigl[ D_{\tilde{i}} D_{\tilde{j}} {}^{(0)}\tilde{n}
 - \Bigl( D_{\tilde{i}} {}^{(0)}\tilde{n} D_{\tilde{j}} \calC
 + D_{\tilde{j}} {}^{(0)}\tilde{n} D_{\tilde{i}} \calC \Bigr) \Bigr]^{T F} 
\notag\\
 & \qquad
  + \frac{3}{a^2 ({}^{(0)}\tilde{N}) e^{2 \calC} f_K ({}^{(0)}\tilde{N})}
 \Bigl[ 1 + \pa_N^2 \log f_K - \bigl(\pa_N \log f_K\bigr)^2 \Bigr] ({}^{(0)}\tilde{N})
 \Bigl( D_{\tilde{i}} {}^{(0)}\tilde{n} D_{\tilde{j}} {}^{(0)}\tilde{n} \Bigr)^{T F} \,.
\label{appV:consis-Aij2}
\end{align}
Comparing Eqs.~(\ref{appV:consis-Aij}) and (\ref{appV:consis-Aij2}),
we see the precise coincidence between the two.

%%%%%%%%%%%%%%%%%%%%%%%%%%%%%%%%%%%%%%%%%%%%%%%%%%%%%%%%%%%%%%%%%%%%%%%%%%
\subsection{Consistency check for $K$}

In the $\calN$ gauge, the evolution equation for $K$ is
\begin{align} 
 \pa_{\tilde{N}} K_\calN &= \frac{3 H (\tilde{N})}{2 \alpha_\calN}
 \bigl(\pa_{\tilde{N}} \phi_\calN\bigr)^2
 - \frac{3}{2 a^2 (\tilde{N}) e^{2\psi_\calN} K_\calN} \Bigl[ R
 - \Bigl( 4 D^2 \psi_\calN + 2 D^{\tilde{i}} \psi_\calN D_{\tilde{i}} \psi_\calN
 \Bigr) \Bigr] \notag\\
 & \qquad
 - \frac{3}{a^2 (\tilde{N}) e^{2\psi_\calN} K_\calN} \left[
 K_\calN D^2 \left( \frac{1}{K_\calN} \right)
 - D^{\tilde{i}} \log K_\calN D_{\tilde{i}} \psi_\calN
 - \frac{1}{2} D^{\tilde{i}} \phi_\calN D_{\tilde{i}} \phi_\calN \right] \,,
\label{appV:N:eq-K}
\end{align}
 and that in the comoving gauge is
\begin{align}
 \pa_N K_c
 &= \frac{3 H (N)}{2 \alpha_c} \bigl(\pa_N f_\phi\bigr)^2 (N)
 - \frac{3}{2 a^2 (N) e^{2\psi_c} f_K (N)} \Bigl[ R
 - \Bigl( 4 D^2 \calC + 2 D^{i} \calC D_{i} \calC \Bigr) \Bigr] \,.
\label{appV:c:eq-K}
\end{align}
 
Again we rewrite the second line inside the brackets of 
Eq.~(\ref{appV:N:eq-K}) as
 \begin{align}  
 (\ma{2nd ~ line}) 
 &= \Bigl[ \bigl(\pa_{\tilde{N}} \log f_K\bigr)^2
 - \pa^2_{\tilde{N}} \log f_K \Bigr] ({}^{(0)}\tilde{N})
 D^{\tilde{i}} {}^{(0)}\tilde{n} D_{\tilde{i}} {}^{(0)}\tilde{n}
 - \pa_{\tilde{N}} \log f_K ({}^{(0)}\tilde{N}) \Bigl( D^2({}^{(0)}\tilde{n})
 + D^{\tilde{i}} {}^{(0)}\tilde{n} D_{\tilde{i}} \calC \Bigr) \,,
\label{appV:2nd-K}
 \end{align}
 where we have used Eq.~(\ref{appV:BG:eq}).

By subtracting Eq.~(\ref{appV:c:eq-K}) from Eq.~(\ref{appV:N:eq-K}),
 we find with the aide of Eqs.~(\ref{appV:N:sol-alpha})
 and (\ref{appV:2nd-K}),
\begin{align} 
 \pa_{\tilde{N}} K \Bigr|^\calN_c
 &= \pa_{\tilde{N}} \Bigl( {}^{(2)}\tilde{n} \pa_N f_K (\tilde{N}) \Bigr)
 + 3 \frac{\pa_{\tilde{N}} \log f_K ({}^{(0)}\tilde{N}) - 2}{a^2 ({}^{(0)}\tilde{N})
 e^{2 \calC} f_K ({}^{(0)}\tilde{N})} \Bigl( D^2({}^{(0)}\tilde{n})
 + D^{\tilde{i}} {}^{(0)}\tilde{n} D_{\tilde{i}} \calC \Bigr) \notag\\
 & \qquad
 + 3 \frac{D^{\tilde{i}} {}^{(0)}\tilde{n} D_{\tilde{i}} {}^{(0)}\tilde{n}}{a^2
 ({}^{(0)}\tilde{N}) e^{2 \calC} f_K ({}^{(0)}\tilde{N})} 
 \left[ 1 + \frac{3}{2} \pa_N \log f_K ({}^{(0)}\tilde{N})
 + \pa^2_{\tilde{N}} \log f_K ({}^{(0)}\tilde{N})
 - \bigl(\pa_{\tilde{N}} \log f_K\bigr)^2 ({}^{(0)}\tilde{N}) \right] \,.
\label{appV:consis-K}
\end{align}

On the other hand, taking the time derivative of Eq.~(\ref{appV:N:sol-K}),
which has been obtained by the gauge transformation,
we obtain
\begin{align} 
 \pa_{\tilde{N}} K \Bigr|^\calN_c
 &= \pa_{\tilde{N}} \Bigl( {}^{(2)}\tilde{n} \pa_N f_K (\tilde{N}) \Bigr)
 + 3 \frac{\pa_{\tilde{N}} \log f_K ({}^{(0)}\tilde{N}) - 2}{a^2 ({}^{(0)}\tilde{N})
 e^{2 \calC} f_K ({}^{(0)}\tilde{N})} \Bigl( D^2({}^{(0)}\tilde{n})
 + D^{\tilde{i}} {}^{(0)}\tilde{n} D_{\tilde{i}} \calC \Bigr) \notag\\
 & \qquad
 + 3 \frac{D^{\tilde{i}} {}^{(0)}\tilde{n} D_{\tilde{i}} {}^{(0)}\tilde{n}}{a^2
 ({}^{(0)}\tilde{N}) e^{2 \calC} f_K ({}^{(0)}\tilde{N})} 
 \left[ 1 + \frac{3}{2} \pa_N \log f_K ({}^{(0)}\tilde{N})
 + \pa^2_{\tilde{N}} \log f_K ({}^{(0)}\tilde{N})
 - \bigl(\pa_{\tilde{N}} \log f_K\bigr)^2 ({}^{(0)}\tilde{N}) \right] \,.
\label{appV:consis-K2}
\end{align}
We see that Eqs.~(\ref{appV:consis-K}) and (\ref{appV:consis-K2})
coincide precisely with each other.

%%%%%%%%%%%%%%%%%%%%%%%%%%%%%%%%%%%%%%%%%%%%%%%%%%%%%%%%%%%%%%%%%%%%%%%%%%
\section{Note on constraints}
\label{app:note}

In general relativity, one cannot freely choose the initial values
 of the metric and matter field variables.
They must be chosen so as to satisfy the Hamiltonian and
momentum constraints. 
In this Appendix we make some comments on these constraint equations
 in the context of the spatial gradient gradient expansion.

Let us consider $\calM$ scalar fields minimally coupled with
Einstein gravity and count the number of degrees of freedom in this system.
In the ADM (Hamiltonian) formalism, the dynamical geometrical 
degrees of freedom are six spatial metric components $(\psi, \gamma_{i j})$
 and their time derivatives $(K, A_{i j})$. 
In addition, we have $2 \calM$ dynamical degrees of freedom associated
 with the scalar fields and their derivatives.
To summarise,
\begin{gather}
 C^\psi : 1 \,, \qquad C_{i j}^\gamma : 5 \,, \qquad C^K : 1 \,, \qquad
 C_{i j}^A : 5 \,, \qquad C_I^\phi : \calM \,, \qquad D_I^\phi : \calM \,,
\end{gather}
that is, we have $12 + 2 \calM$ ($=6+6+\calM+\calM$) degrees of freedom.
The Hamiltonian and momentum constraint equations give four constraints 
among them. And there are also four gauge degrees of freedom.
Thus there are $4 + 2 \calM$ ($= 12+ 2 \calM -(4 + 4)$) independent
 physical degrees of freedom left, which represent $2 \times 2$
 gravitational wave degrees of freedom and $\calM \times 2$ scalar field
degrees of freedom. 

Let us explicitly check how the above counting works  
 in the $\calN$ gauge, where we take the uniform $\calN$ slicing and
 the time-slice-orthogonal threading.
The $2\calM$ scalar field degrees are represented by $C_I^\phi$ and $D_I^\phi$.
The gravitational wave degrees of freedom are contained in
 among 5 degrees of freedom in $C_{i j}^\gamma$ and $C_{i j}^A$, respectively,
2 of each describe the gravitational degrees of freedom.
The remaining 3 degrees of freedom in $C_{ij}^\gamma$ are fixed by
the threading condition $\beta^i = 0$ together with purely
spatial coordinate transformation degrees of freedom (see e.g.,
Eq.~(\ref{cond:gamma-flat}) in the example of a canonical scalar field),
 and those 3 in $C_{ij}^A$ are fixed by the momentum constraint~(\ref{sb:eq-CAij}).
The Hamiltonian constraint is used to determine $C^K$.

Now we are left with $C^\psi$. This may be regarded as the
remaining gauge degree of freedom in the $\calN$ slicing,
since the slicing condition implies $\pa_t \psi = 0$.
But once the scalar field configuration on the initial slice 
is fixed, it should not be freely specifiable, since
if it were it would contradict with the total number of
the physical degrees of freedom counted above for the general case.
In subsection~\ref{sapp:Ham}, we find there is indeed
a constraint equation that must be satisfied by $C^\psi$.
Thus we recover the total number of the true physical degrees
of freedom correctly. 

Recently the validity of the $\delta N$ formalism
was studied by Sugiyama et al.~\cite{Sugiyama:2012tj},
and they found the violation of the momentum constraint 
 in a non-slow-roll inflation case.
In subsection~\ref{sapp:Mom}, we show how the consistency with
the momentum constraint is recovered in the gradient expansion.

%%%%%%%%%%%%%%%%%%%%%%%%%%%%%%%%%%%%%%%%%%%%%%%%%%%%%%%%%%%%%%%%%%%%%%%%%%
\subsection{``Hidden'' Hamiltonian constraint}
\label{sapp:Ham}

In general the Hamiltonian constraint gives a non-trivial relation
 among the initial values of the metric components, their time
derivatives and the matter field variables.
However, in Sec.~\ref{sec:sbrid},
we have used the Hamiltonian constraint to solve for $K$.
Thus it seems it will not give any more constraint.
Then where is a constraint corresponding to the Hamiltonian constraint? 
The answer is that it is hided in the evolution equation for $K$.

In order to see it explicitly, let us consider the canonical 
scalar field model discussed in Sec.~\ref{sec:sbrid}.
We first solve the evolution equation for $K$, Eq.~(\ref{N:eq-K}), 
and then compare its solution with the solution obtained 
form the Hamiltonian constraint.

Let us consider the evolution equation for ${}^{(2)}K$
 in the $\calN$ gauge. Using the leading order solution and ${}^{(2)}\phi$
given by Eq.~(\ref{sb:sol-phi-2}), we find
\begin{align}
 \pa_N \left( \frac{{}^{(2)}K}{\sqrt{3 {}^{(0)}V}} \right)
 &= m_I \left( D_I^\phi e^{3 (N - N_0)}
 + \frac{S_I^\phi}{a^3 e^{2 C^\psi}} \int d N' \frac{a}{{}^{(0)}V} \right) \notag\\
 & \qquad - \frac{1}{2 a^2 e^{2 C^\psi} {}^{(0)}V} \Bigl[ R
 - \Bigl( 4 D^2 C^\psi + 2 D^i C^\psi D_i C^\psi \Bigr)
 - D^i C^\phi_I D_i C^\phi_I \Bigr] \,.
\end{align}
This is easily integrated to give
\begin{align}
 \frac{{}^{(2)}K_\ma{tmp}(N)}{\sqrt{3 {}^{(0)}V}} 
 &= \frac{{}^{(2)}K_\ma{tmp}(N_0)}{\sqrt{3 {}^{(0)}{V}_0}}
 + \frac{1}{3} m_I D_I^\phi \Bigl[ e^{3 (N - N_0)} - 1 \Bigr]
 + \frac{m_I S_I^\phi I_\phi (N)}{a^2 e^{2 C^\psi}{}^{(0)}V} \notag\\
 & \qquad
 - \frac{ R - \Bigl( 4 D^2 C^\psi + 2 D^i C^\psi D_i C^\psi \Bigr)
 - D^i C^\phi_I D_i C^\phi_I}{2 a^2 e^{2 C^\psi} {}^{(0)}V}
 \left( a^2 {}^{(0)}V \int d N' \frac{1}{a^2 {}^{(0)}V} \right) \,,
\label{app:N:sol-K}
\end{align}
where the suffix tmp indicates that it is obtained from the
temporal evolution equation.

Now let us carefully examine the solution~(\ref{app:N:sol-K})
in comparison with that obtained in the text, Eq.~(\ref{sb:sol-K}).
First we compare the initial values. This determines
 the initial value of ${}^{(2)}K_\ma{tmp}$ as
\begin{align}
 \frac{{}^{(2)}K_\ma{tmp}(N_0)}{\sqrt{3 {}^{(0)}{V}_0}}
 &= \frac{1}{6} m_I D_I^\phi - \frac{1}{4 a_0^2 e^{2 C^\psi} {}^{(0)}{V}_0}
 \Bigl[ R - \Bigl( 4 D^2 C^\psi + 2 D^i C^\psi D_i C^\psi \Bigr)
 + D^i C^\phi_I D_i C^\phi_I \Bigr] \,.
\label{app:N:sol-K-ini}
\end{align} 
Next we subtract the solution~(\ref{app:N:sol-K}) together with thus
obtained initial data~(\ref{app:N:sol-K-ini}) 
from the solution~(\ref{sb:sol-K}),
\begin{align}
 \frac{{}^{(2)}K - {}^{(2)}K_\ma{tmp}}{\sqrt{3 {}^{(0)}V}}
 &= \frac{m_I S_I^\phi}{a^2 e^{2 C^\psi} {}^{(0)}V} \left(
 \frac{1}{6} \frac{{}^{(0)}V}{a} \int^N_{N_0} dN' \frac{a}{{}^{(0)}V}
 - \frac{1}{2} I_\phi (N) \right) \notag\\
 & \qquad
 + \frac{1}{4 e^{2 C^\psi}} \left( \frac{1}{a_0^2 {}^{(0)}{V}_0} - 
\frac{1}{a^2 {}^{(0)}V}
 \right) \Bigl[ R - \Bigl(4 D^2 C^\psi + 2 D^i C^\psi D_i C^\psi\Bigr)
 + D^i C^\phi_I D_i C^\phi_I \Bigr] \notag\\
 & \qquad
 + \frac{ R - \Bigl( 4 D^2 C^\psi + 2 D^i C^\psi D_i C^\psi \Bigr)
 - D^i C^\phi_I D_i C^\phi_I}{2 a^2 e^{2 C^\psi} {}^{(0)}V}
 \left( a^2 {}^{(0)}V \int d N' \frac{1}{a^2 {}^{(0)}V} \right) \,.
\end{align}
After integration parts and some manipulations, we obtain
\begin{align}
 \frac{{}^{(2)}K - {}^{(2)}K_\ma{tmp}}{\sqrt{3{}^{(0)}V}}
 &= \frac{1}{2 a^2 e^{2 C^\psi} {}^{(0)}V} \left\{ \frac{m_I}{e^{C^\psi}}
 D_i\left( e^{C^\psi} D^i C_I^\phi \right)
 - \frac{M^2 + 6}{6} D^i C^\phi_I D_i C^\phi_I \right. \notag\\
 & \qquad \qquad \qquad \qquad \qquad \left.
 + \frac{M^2}{6} \Bigl[ R - \Bigl( 4 D^2 C^\psi + 2 D^i C^\psi D_i C^\psi \Bigr)
 \Bigr] \right\} \left( a^2 {}^{(0)}V \int d N' \frac{1}{a^2{}^{(0)}V} \right)
 \,.
\end{align}
The right-hand side of the above equation must vanish at all times.
This demands that the term inside the square brackets should vanish,
 \begin{align}
 \frac{m_I}{e^{C^\psi}}D_i\left( e^{C^\psi} D^i C_I^\phi \right)
 - \frac{M^2 + 6}{6} D^i C^\phi_I D_i C^\phi_I 
 + \frac{M^2}{6} \Bigl[ R - \Bigl(4 D^2 C^\psi + 2 D^i C^\psi D_i C^\psi
 \Bigr) \Bigr] = 0 \,.  
 \end{align}
Apparently this gives a non-trivial constraint among initial data,
$C^\psi$ and $C^\phi_I$. This is the ``hidden'' constraint corresponding 
to the Hamiltonian constraint.

%%%%%%%%%%%%%%%%%%%%%%%%%%%%%%%%%%%%%%%%%%%%%%%%%%%%%%%%%%%%%%%%%%%%%%%%%%
\subsection{Consistency with the momentum constraint}
\label{sapp:Mom}

Below we show that the momentum constraint is automatically satisfied
 if the Hamiltonian constraint as well as the scalar field equation
is satisfied. In the case of slow-roll inflation,
 the leading order Hamiltonian constraint is sufficient to show it.
For general (non-slow-roll) inflation, one needs
 the next-leading order Hamiltonian constraint.

\subsubsection{Slow-roll case}
\label{sapp:slow}

Under the slow-roll approximation, the Hamiltonian
and momentum constraints reduce respectively to
\begin{align}
 \frac{2}{3} K^2
 &= - 2 P \bigl( X^{I J} (\phi^K), \phi^L \bigr) \,,
\label{appn:Ham-slow} \\
 \frac{2}{3} \pa_i K
 &= \frac{2 K}{3} P_{(I J)} \pa_N \phi^I \pa_i \phi^J \,, 
\label{appn:Mom-slow}
\end{align}
 and the scalar field equation becomes
 \begin{align}
 - \frac{K^2}{3} P_{(I J)} \pa_N \phi^J - \frac{1}{2} P_I = 0 \,.
\label{appn:eq-phi-slow}
 \end{align}

Now we multiply both sides of the momentum constraint~(\ref{appn:Mom-slow})
 by $2K$, and substitute the scalar field equation~(\ref{appn:eq-phi-slow})
 into the right-hand side of it. This gives
\begin{align}
 \ma{(left-hand side)} 
&= \frac{4}{3} K \pa_i K = \pa_i \left( \frac{2}{3} K^2 \right)
 \,, \\
 \ma{(right-hand side)} 
 &= - 2 P_I \pa_i \phi^I = \pa_i \bigl(- 2 P\bigr) \,.
\end{align}
One can easily see that these are merely a spatial derivative of 
the Hamiltonian constraint~(\ref{appn:Ham-slow}).
Thus we conclude that, at leading order in gradient expansion,
the momentum constraint is automatically satisfied if
the Hamiltonian constraint and the scalar field equations
are satisfied.

\subsubsection{Non slow-roll case}
\label{sapp:nonslow}

Here we show that in general the leading order momentum constraint is
 automatically satisfied once the Hamiltonian constraint and
the energy conservation equations are satisfied
 to next-leading order in gradient expansion. 

From Eqs.~(\ref{basic-H}) and (\ref{basic-M}) with $A_{ij}=\calO(\epsilon^2)$, 
the Hamiltonian and momentum constraint equations are 
\begin{align}
 \frac{1}{a^2 e^{2\psi}} \Bigl[ R - \Bigl( 4 D^2 \psi+2D^i \psi D_i \psi \Bigr)
 \Bigr] + \frac{2}{3} K^2 &= 2E +\calO(\epsilon^4)\,, \label{sbb:nonslow-H}\\ 
 \frac{2}{3} \pa_i K &= J_i +\calO(\epsilon^3)\,.
\label{sbb:nonslow-M}
\end{align}
The energy conservation law, $n^\nu\nabla_\nu T^{\mu \nu} = 0$, 
 in the $\calN$ gauge is
\begin{align}
 \pa_N E
 = 3 \bigl(E + P\bigr) + \frac{3 K}{a^2 e^{3 \psi}} D^i \left(
 \frac{e^\psi}{K^2} J_i \right)
 + \frac{2}{a^2 e^{2 \psi}} P_{(I J)} D^i \phi^I \pa_j \phi^J \,.
\label{appn:eq-E}
\end{align}
%%%%%%%%%%%%%% October 15, 2012

Now let us multiply the evolution equation for $K$, Eq.~(\ref{N:eq-K}),
by $2 K/3$ and subtract Eq.~(\ref{appn:eq-E}) from it. This gives
\begin{align}
 &\pa_N \left( \frac{1}{3} K^2 - E \right)
\notag\\ 
&\quad
= - \frac{1}{a^2 e^{2 \psi}} \Bigl[ R
 - \Bigl( 4 D^2 \psi +2 D^i \psi D_i \psi \Bigr) \Bigr]
 - \frac{2 K}{a^2 e^{2 \psi}} \left[ D^2 \left( \frac{1}{K} \right)
 + D^i \left( \frac{1}{K} \right) D_i \psi \right] 
 - \frac{3 K}{a^2 e^{3 \psi}} D^i \left( \frac{e^\psi}{K^2} J_i \right) \,.
\end{align}
Substituting the Hamiltonian constraint~(\ref{sbb:nonslow-H}) 
into the left-hand side of the above equation, we find
\begin{align}  
 \pa_N \left( \frac{1}{3} K^2 - E \right)
 = - \pa_N \left\{ \frac{1}{2 a^2 e^{2 \psi}} \Bigl[ R
 - \Bigl(4 D^2 \psi +2 D^i \psi D_i \psi \Bigr) \Bigr] \right\}
 = - \frac{1}{a^2 e^{2 \psi}} \Bigl[ R
 - \Bigl(4 D^2 \psi +2 D^i \psi D_i \psi \Bigr) \Bigr] \,.
\end{align}
Comparing the above two equations, we obtain 
\begin{align}
 0
 &= - \frac{2}{3} e^\psi \left[ D^2 \left( \frac{1}{K} \right)
 + D^i \left( \frac{1}{K} \right) D_i \psi \right] 
 - D^i \left( \frac{e^\psi}{K^2} J_i \right) \notag\\
 &= D^i \left[ \frac{e^\psi}{K^2} \left( \frac{2}{3} \pa_i K
 - J_i \right) \right] \,.
\end{align}
Again one can see that the leading order
momentum constraint~(\ref{sbb:nonslow-M}) holds automatically if the
next-to-leading order Hamiltonian constraint~(\ref{sbb:nonslow-H}) holds.

What does this mean?
In the gradient expansion, we have assumed $A_{i j}$ does not contribute to
 the leading order dynamics. This corresponds to neglecting 
the adiabatic decaying mode in linear theory.
In a single field model, it is well known this decaying mode appears
only at the next-leading order both in linear theory~\cite{Leach:2001zf}
 and in non-linear theory~\cite{Takamizu:2010xy,Takamizu:2010je}. 
We also mention the work~\cite{Nalson:2011gc} in which
they discussed the behavior of decaying modes in different choices of gauge.
The absence of the decaying mode at leading order in gradient expansion
should hold also in the case of multi-field inflation, at least as long
as the background homogeneous solution is stable against 
a homogeneous but anisotropic perturbation.

To summarise, the point is that the initial data for the scalar field
and its time derivatives are not freely specifiable in general,
 and in particular we need to take into account the next-leading order
terms in the Hamiltonian constraint for the non-slow-roll case.
Once we take into account the next-leading order Hamiltonian constraint, 
 the leading order momentum constraint is automatically satisfied.
As for the next-to-leading order momentum constraint, it constrains
the initial value of $A_{ij}$ as in Eq.~(\ref{sb:eq-CAij}).

%%%%%%%%%%%%%%%%%%%%%%%%%%%%%%%%%%%%%%%%%%%%%%%%%%%%%%%%%%%%%%%%%%%%%%%%%%


\begin{thebibliography}{100}

\bibitem{Komatsu:2010fb}
WMAP, E.~Komatsu {\em et~al.},
\newblock Astrophys. J. Suppl. {\bf 192}, 18 (2011), arXiv:1001.4538.
%%CITATION = 1001.4538;%%

%\cite{:2006uk}
\bibitem{Planck:2006uk}
    [Planck Collaboration], 
 %``Planck: The scientific programme,''
  arXiv:astro-ph/0604069.

\bibitem{CQG-focus-NG}
M.~Sasaki and D.~Wands,
\newblock Classical and Quantum Gravity {\bf 27}, 120301 (2010).

\bibitem{Lifshitz:1963ps}
E.~M. Lifshitz and I.~M. Khalatnikov,
\newblock Adv. Phys. {\bf 12}, 185 (1963).
%%CITATION = ADPHA,12,185;%%

\bibitem{Belinsky:1982pk}
V.~a. Belinsky, I.~m. Khalatnikov, and E.~m. Lifshitz,
\newblock Adv. Phys. {\bf 31}, 639 (1982).
%%CITATION = ADPHA,31,639;%%

\bibitem{Starobinsky:1986fxa}
A.~A. Starobinsky,
\newblock JETP Lett. {\bf 42}, 152 (1985).
%%CITATION = JTPLA,42,152;%%

\bibitem{Bardeen:1980kt}
J.~M. Bardeen,
\newblock Phys. Rev. {\bf D22}, 1882 (1980).
%%CITATION = PHRVA,D22,1882;%%

\bibitem{Salopek:1990jq}
D.~S. Salopek and J.~R. Bond,
\newblock Phys. Rev. {\bf D42}, 3936 (1990).
%%CITATION = PHRVA,D42,3936;%%

\bibitem{Deruelle:1994iz}
N.~Deruelle and D.~Langlois,
\newblock Phys. Rev. {\bf D52}, 2007 (1995), arXiv:gr-qc/9411040.
%%CITATION = GR-QC/9411040;%%

\bibitem{Nambu:1994hu}
Y.~Nambu and A.~Taruya,
\newblock Class. Quant. Grav. {\bf 13}, 705 (1996), arXiv:astro-ph/9411013.
%%CITATION = ASTRO-PH/9411013;%%

\bibitem{Sasaki:1995aw}
M.~Sasaki and E.~D. Stewart,
\newblock Prog. Theor. Phys. {\bf 95}, 71 (1996), arXiv:astro-ph/9507001.
%%CITATION = ASTRO-PH/9507001;%%

\bibitem{Sasaki:1998ug}
M.~Sasaki and T.~Tanaka,
\newblock Prog. Theor. Phys. {\bf 99}, 763 (1998), arXiv:gr-qc/9801017.
%%CITATION = GR-QC/9801017;%%

\bibitem{Shibata:1999zs}
M.~Shibata and M.~Sasaki,
\newblock Phys. Rev. {\bf D60}, 084002 (1999), arXiv:gr-qc/9905064.
%%CITATION = GR-QC/9905064;%%

\bibitem{Kodama:1997qw}
  H.~Kodama and T.~Hamazaki, \newblock
  %``Evolution of cosmological perturbations in the long wavelength limit,''
  Phys.\ Rev.\  D {\bf 57}, 7177 (1998), arXiv:gr-qc/9712045.

\bibitem{Hamazaki:2008mh}
  T.~Hamazaki, \newblock
 %``Long wavelength limit of evolution of nonlinear cosmological
 % perturbations,''
  Phys.\ Rev.\  D {\bf 78}, 103513 (2008), arXiv:0811.2366.


\bibitem{Wands:2000dp}
D.~Wands, K.~A. Malik, D.~H. Lyth, and A.~R. Liddle,
\newblock Phys. Rev. {\bf D62}, 043527 (2000), arXiv:astro-ph/0003278.
%%CITATION = ASTRO-PH/0003278;%%

\bibitem{Lyth:2004gb}
D.~H. Lyth, K.~A. Malik, and M.~Sasaki,
\newblock JCAP {\bf 0505}, 004 (2005), arXiv:astro-ph/0411220.
%%CITATION = ASTRO-PH/0411220;%%

\bibitem{Lyth:2005du}
D.~H. Lyth and Y.~Rodriguez,
\newblock Phys. Rev. {\bf D71}, 123508 (2005), arXiv:astro-ph/0502578.
%%CITATION = ASTRO-PH/0502578;%%

\bibitem{Seery:2005gb}
  D.~Seery and J.~E.~Lidsey,\newblock
 %``Primordial non-gaussianities from multiple-field inflation,''
  JCAP {\bf 0509}, 011 (2005), arXiv:astro-ph/0506056.


%\cite{Tanaka:2006zp}
\bibitem{Tanaka:2006zp} 
  Y.~Tanaka and M.~Sasaki,
  %``Gradient expansion approach to nonlinear superhorizon perturbations,''
  Prog.\ Theor.\ Phys.\  {\bf 117}, 633 (2007), arXiv:gr-qc/0612191.
  %%CITATION = GR-QC/0612191;%%

%\cite{Tanaka:2007gh}
\bibitem{Tanaka:2007gh} 
  Y.~Tanaka and M.~Sasaki,
  %``Gradient expansion approach to nonlinear superhorizon perturbations. II. A Single scalar field,''
  Prog.\ Theor.\ Phys.\  {\bf 118}, 455 (2007), arXiv:0706.0678.
  %%CITATION = ARXIV:0706.0678;%%

\bibitem{Yokoyama:2007uu}
S.~Yokoyama, T.~Suyama, and T.~Tanaka,
\newblock JCAP {\bf 0707}, 013 (2007), arXiv:0705.3178.
%%CITATION = 0705.3178;%%

\bibitem{Sasaki:2008uc}
  M.~Sasaki,\newblock
 %``Multi-brid inflation and non-Gaussianity,''
  Prog.\ Theor.\ Phys.\  {\bf 120}, 159 (2008), arXiv:0805.0974.

\bibitem{Yokoyama:2007dw}
S.~Yokoyama, T.~Suyama, and T.~Tanaka,
\newblock Phys. Rev. {\bf D77}, 083511 (2008), arXiv:0711.2920.
%%CITATION = 0711.2920;%%

\bibitem{Weinberg:2008nf}
S.~Weinberg,
\newblock Phys.Rev. {\bf D78}, 123521 (2008), arXiv:0808.2909.
%%CITATION = ARXIV:0808.2909;%%

\bibitem{Weinberg:2008si}
S.~Weinberg,
\newblock Phys. Rev. {\bf D79}, 043504 (2009), arXiv:0810.2831.
%%CITATION = 0810.2831;%%

\bibitem{Takamizu:2008ra}
Y.-i. Takamizu and S.~Mukohyama,
\newblock JCAP {\bf 0901}, 013 (2009), arXiv:0810.0746.
%%CITATION = 0810.0746;%%

\bibitem{Takamizu:2010xy}
Y.-i. Takamizu, S.~Mukohyama, M.~Sasaki, and Y.~Tanaka,
\newblock JCAP {\bf 1006}, 019 (2010), arXiv:1004.1870.
%%CITATION = 1004.1870;%%

\bibitem{Takamizu:2010je}
Y.-i. Takamizu and J.~Yokoyama,
\newblock Phys.Rev. {\bf D83}, 043504 (2011), arXiv:1011.4566.
%%CITATION = ARXIV:1011.4566;%%

\bibitem{Sugiyama:2012tj}
N.~S. Sugiyama, E.~Komatsu, and T.~Futamase,
\newblock (2012), arXiv:1208.1073.
%%CITATION = ARXIV:1208.1073;%%

\bibitem{Leach:2001zf}
S.~M. Leach, M.~Sasaki, D.~Wands, and A.~R. Liddle,
\newblock Phys.Rev. {\bf D64}, 023512 (2001), arXiv:astro-ph/0101406.
%%CITATION = ASTRO-PH/0101406;%%

\bibitem{Seto:1999jc}
O.~Seto, J.~Yokoyama, and H.~Kodama,
\newblock Phys. Rev. {\bf D61}, 103504 (2000), arXiv:astro-ph/9911119.
%%CITATION = ASTRO-PH/9911119;%%

\bibitem{Namjoo:2012aa} 
M.~H.~Namjoo, H.~Firouzjahi and M.~Sasaki,
\newblock (2012), arXiv:1210.3692.
%%CITATION = ARXIV:1210.3692;%%

\bibitem{Choi:2007su}
K.-Y. Choi, L.~M. Hall, and C.~van~de Bruck,
\newblock JCAP {\bf 0702}, 029 (2007), arXiv:astro-ph/0701247.
%%CITATION = ASTRO-PH/0701247;%%

\bibitem{Byrnes:2009qy}
C.~T. Byrnes and G.~Tasinato,
\newblock JCAP {\bf 0908}, 016 (2009), arXiv:0906.0767.
%%CITATION = ARXIV:0906.0767;%%

%\cite{Langlois:2008qf}
\bibitem{Langlois:2008qf} 
  D.~Langlois, S.~Renaux-Petel, D.~A.~Steer and T.~Tanaka, 
\newblock Phys.\ Rev.\ D {\bf 78}, 063523 (2008), arXiv:0806.0336.
  %%CITATION = ARXIV:0806.0336;%%

%\cite{Arroja:2008yy}
\bibitem{Arroja:2008yy} 
  F.~Arroja, S.~Mizuno and K.~Koyama,
\newblock JCAP {\bf 0808}, 015 (2008), arXiv:0806.0619.
  %%CITATION = ARXIV:0806.0619;%%

%\cite{Emery:2012sm}
\bibitem{Emery:2012sm} 
  J.~Emery, G.~Tasinato and D.~Wands, 
\newblock JCAP {\bf 1208}, 005 (2012), arXiv:1203.6625.
  %%CITATION = ARXIV:1203.6625;%%

%\cite{Kidani:2012jp}
\bibitem{Kidani:2012jp} 
  T.~Kidani, K.~Koyama and S.~Mizuno, 
  \newblock arXiv:1207.4410.
  %%CITATION = ARXIV:1207.4410;%%

%\cite{Kodama:1985bj}   
\bibitem{Kodama:1985bj}
  H.~Kodama and M.~Sasaki, 
 %``Cosmological Perturbation Theory,''
  Prog.\ Theor.\ Phys.\ Suppl.\  {\bf 78} 1 (1984).

%\cite{Nalson:2011gc}
\bibitem{Nalson:2011gc} 
  E.~Nalson, A.~J.~Christopherson, I.~Huston and K.~A.~Malik, 
\newblock arXiv:1111.6940.
  %%CITATION = ARXIV:1111.6940;%%
\end{thebibliography}
\end{document}